\pdfoutput=1
\documentclass[fleqn,usenatbib]{mnras}
\usepackage{newtxtext,newtxmath}
\usepackage[T1]{fontenc}
\usepackage{ae,aecompl}

\usepackage{graphicx}	% Including figure files
\usepackage{amsmath}	% Advanced maths commands
\usepackage{amssymb}	% Extra maths symbols
\usepackage{booktabs}

\title[HD 209458b in New Light]{HD 209458b in New Light: Evidence of Nitrogen Chemistry, Patchy Clouds and Sub-Solar Water}

\author[MacDonald \& Madhusudhan]{
Ryan J. MacDonald$^{1}$\thanks{Email: r.macdonald@ast.cam.ac.uk}
\& Nikku Madhusudhan$^{1}$\thanks{Email: nmadhu@ast.cam.ac.uk}
\\
$^{1}$Institute of Astronomy, University of Cambridge, Madingley Road, Cambridge, CB3 0HA, UK
}

\date{Accepted 2017 March 30. Received 2017 March 29; in original form 2017 January 4}

\pubyear{2017}

\begin{document}
\label{firstpage}
\pagerange{\pageref{firstpage}--\pageref{lastpage}}
\maketitle

% Abstract of the paper
\begin{abstract}
Interpretations of exoplanetary transmission spectra have been undermined by apparent obscuration due to clouds/hazes. Debate rages on whether weak H$_2$O features seen in exoplanet spectra are due to clouds or inherently depleted oxygen. Assertions of solar H$_2$O abundances have relied on making a priori model assumptions, e.g. chemical/radiative equilibrium. In this work, we attempt to address this problem with a new retrieval paradigm for transmission spectra. We introduce POSEIDON, a two-dimensional atmospheric retrieval algorithm including generalised inhomogeneous clouds. We demonstrate that this prescription allows one to break vital degeneracies between clouds and prominent molecular abundances. We apply POSEIDON to the best transmission spectrum presently available, for the hot Jupiter HD~209458b, uncovering new insights into its atmosphere at the day-night terminator. We extensively explore the parameter space with an unprecedented 10$^8$ models, spanning the continuum from fully cloudy to cloud-free atmospheres, in a fully Bayesian retrieval framework. We report the first detection of nitrogen chemistry (NH$_3$ and/or HCN) in an exoplanet atmosphere at 3.7-7.7$\sigma$ confidence, non-uniform cloud coverage at 4.5-5.4$\sigma$, high-altitude hazes at $>$3$\sigma$, and sub-solar H$_2$O at $\gtrsim$3-5$\sigma$, depending on the assumed cloud distribution. We detect NH$_3$ at 3.3$\sigma$ and 4.9$\sigma$ for fully cloudy and cloud-free scenarios, respectively. For the model with the highest Bayesian evidence, we constrain H$_2$O at 5-15 ppm (0.01-0.03$\times$ solar) and NH$_3$ at $0.01-2.7$ ppm, strongly suggesting disequilibrium chemistry and cautioning against equilibrium assumptions. Our results herald new promise for retrieving cloudy atmospheres using high-precision HST and JWST spectra. 
\end{abstract}


\section{Introduction}\label{Intro}

We stand at the precipice of a new age -- one where the vision of characterising exoplanets in exquisite detail is rapidly being realised. Tremendous progress has been made in recent years towards observing various aspects of exoplanetary atmospheres, including chemical signatures \citep{Snellen2010,Deming2013}, temperature profiles \citep{Haynes2015,Line2016b}, circulation patterns \citep{Stevenson2014}, clouds/hazes \citep{Sing2016}, and escape processes \citep{Ehrenreich2015}; for recent reviews see e.g. \citet{Madhusudhan2014a, Crossfield2015, Madhusudhan2016a}. Using state-of-the-art atmospheric retrieval techniques, it is now also possible to use spectroscopic observations to extract precise constraints on the chemical abundances. Such constraints are just beginning to provide tantalising clues to planetary formation and migration pathways \citep[e.g.][]{Oberg2011, Madhusudhan2014b, Mordasini2016}. While a variety of observations have been used to study exoplanetary atmospheres, the majority have focused on transiting hot Jupiters ($T\sim 800-3000$ K) whose extended atmospheres and favourable geometry make them especially amenable to transit spectroscopy. 

The most observed molecule in exoplanetary atmospheres to date is H$_2$O. In recent years, the Hubble Space Telescope's (HST) Wide Field Camera 3 (WFC3) has enabled robust detections of H$_2$O in numerous exoplanetary transmission spectra \citep[e.g.][]{Deming2013}. However, in almost all cases the amplitudes of H$_2$O absorption features are significantly lower than those expected of a cloud-free solar-composition atmosphere -- typically $\sim$2 scale heights \citep{Deming2013,Kreidberg2015,Sing2016} instead of $\sim$5-10 \citep{Madhusudhan2014a}. Taken at face value, this could imply a plethora of atmospheres inherently depleted in oxygen \citep{Madhusudhan2014c}. Alternatively, they may be explained by invoking a high altitude ($P<$1 mbar) opaque cloud deck \citep{Deming2013} or uniform-in-altitude grey opacity, such as haze particles \citep{Pont2013}. Given the increasing number of low-amplitude or even flat spectra observed \citep[e.g][]{Kreidberg2014a, Knutson2014b, Knutson2014c, Ehrenreich2014a, Sing2016} the consideration of clouds has been elevated to the forefront of transmission spectroscopy.

The fundamental issue with deriving chemical abundances in cloudy exoplanetary atmospheres lies in innate degeneracies between clouds and chemistry. A wide range of solutions exist, spanning high-altitude clouds with concealed solar abundances and low-altitude, or non-existent, clouds with sub-solar abundances. This naturally leads to extremely loose constraints consistent with the full range from sub-solar through super-solar abundances \citep[e.g.,][]{Benneke2015}. It is thus clear that clouds pose an existential challenge to robustly estimating chemical abundances. Most efforts to retrieve atmospheric properties of cloudy atmospheres have employed one-dimensional models -- i.e. homogeneous terminator cloud coverage. This is despite predictions from Global Circulation Models (GCMs) that large temperature contrasts of many hundreds of K may fuel a prominence of \emph{partially cloudy} terminators on tidally locked hot Jupiters \citep{Parmentier2016}. The effect of partial clouds on transmission spectra retrieval has recently been examined by \citet{Line2016a}. 

Here, we offer a potential solution to the problem of interpreting transmission spectra of cloudy exoplanets. We introduce POSEIDON, a new atmospheric retrieval algorithm that includes generalised two-dimensional inhomogeneous cloud distributions. By not assuming global cloud coverage across the terminator, regions without clouds are sampled during transmission -- effectively allowing one to `peer below' the clouds and break many of the degeneracies between clouds and chemical abundances. The method we propose enables the \emph{simultaneous} retrieval of cloud/haze properties along with precise molecular abundance constraints. 

We demonstrate our new retrieval paradigm on the best transmission spectrum available, namely that of the hot Jupiter HD 209458b. As the first transiting exoplanet \citep{Charbonneau2000}, HD 209458b ignited the fledgling field of exoplanetary atmospheres. Serving as the prototypical target for atmospheric characterisation, it was the first exoplanet observed to contain Na \citep{Charbonneau2002}. Various carbon and oxygen-rich molecular species have been claimed in its atmosphere, including $\mathrm{H_2 O}$, $\mathrm{CH_4}$, $\mathrm{CO}$ and $\mathrm{CO_2}$ \citep{Barman2007, Swain2009, Madhusudhan2009,Snellen2010,Deming2013}. Of these detections, $\mathrm{H_2 O}$ has been robustly verified by WFC3 spectroscopy \citep{Deming2013} and $\mathrm{CO}$ by high resolution Doppler spectroscopy \citep{Snellen2010}. 

Constraining abundances of chemical species became possible with the invention of atmospheric retrieval for exoplanets \citep{Madhusudhan2009,Madhusudhan2011}. Retrieval techniques allow atmospheric properties to be derived directly from observational data in a statistically robust manner. Retrieval is now a mature field, with a wide range of codes developed and deployed to analyse atmospheres in transmission \citep{Benneke2012,Benneke2013,Benneke2015, Waldmann2015a, Line2016a}, thermal emission \citep{Lee2012,Barstow2013,Line2013,Line2016b,Waldmann2015b} and reflected light \citep{Lee2013,Lupu2016a,Lavie2016}. The most robust inferences of the molecular abundances to date have been derived from near-infrared spectra obtained by HST WFC3 in the $\sim 1.1-1.7 \, \micron$ range which contains strong spectral features due to $\mathrm{H_2 O}$, $\mathrm{CH_4}$, $\mathrm{NH_3}$ and $\mathrm{HCN}$.

The WFC3 transmission spectrum of HD 209458b \citep{Deming2013} was first retrieved by \citet{Madhusudhan2014c}. They reported a water abundance of $\sim$0.01-0.05$\times$ the prediction for a solar abundance atmosphere ($\sim$ 5$\times 10^{-3}$), despite a super-solar stellar metallicity of [O/H] = $0.092 \pm 0.036$ \citep{Brewer2016}. Such low inferred $\mathrm{H_2 O}$ abundances may provide interesting constraints on planetary formation and migration \citep{Madhusudhan2014b}. However, this retrieval assumed a cloud-free atmosphere. \citet{Benneke2015} revisited this spectrum with a retrieval including clouds, albeit with assumed radiative-convective equilibrium along with a priori C-N-O chemistry, and inferred a composition consistent with solar abundances (also see section~\ref{section:discussion}). This view was further reinforced by \citet{Sing2016}, who used transmission spectra of ten hot Jupiters to claim that clouds and hazes, not sub-solar H$_2$O, sufficiently explain the spectra. However, this work only considered a small grid of forward models with chemical equilibrium assumed, rather than a retrieval. In another effort, \citet{Tsiaras2016a} also detected H$_2$O but were unable to robustly constrain its abundance. Suggestions of solar $\mathrm{H_2 O}$ abundances have since been called into question. \citet{Barstow2016} performed retrievals of the \citet{Sing2016} datasets and found that nine of their ten planets possess sub-solar water abundances once the assumption of chemical equilibrium is relaxed -- with HD 209458b the driest at $\sim 0.01-0.02 \times$ solar. These competing lines of evidence leaves the question of sub-solar water abundances unsettled.

In this work, we examine these differing conclusions through the application of our two-dimensional retrieval code, POSEIDON. Using a state-of-the-art nested sampling algorithm, we extensively explore the model parameter space in a fully Bayesian retrieval framework with $>$$10^{8}$ model spectra. Unlike the retrievals of \citet{Madhusudhan2014c} and \citet{Barstow2016}, we additionally include the nitrogen-bearing molecules $\mathrm{NH_3}$ and $\mathrm{HCN}$. In what follows, we introduce our retrieval framework in \S\ref{section:methods}. We validate POSEIDON using a simulated dataset in \S\ref{section:validation}. We retrieve the atmospheric properties of HD~209458b in \S\ref{section:results}. Finally, in \S\ref{section:discussion} we summarise our results and discuss the implications. 

\section{Atmospheric Retrieval with Inhomogeneous Clouds}\label{section:methods}

Here we introduce POSEIDION, our modelling and retrieval framework for transmission spectra. Extracting atmospheric properties from an observed spectrum involves two components: i) a parametric \emph{forward model}; ii) a \emph{statistical retrieval} algorithm to sample the model parameter space. Typical forward models assume one-dimensional geometry, i.e. average temperature, composition and cloud properties across the terminator. In this work we generalise the formulation to account for azimuthally inhomogeneous cloud properties.

\begin{figure*}
	\includegraphics[width=\linewidth,  trim={0.0cm 0.0cm 0.0cm 0.1cm}]{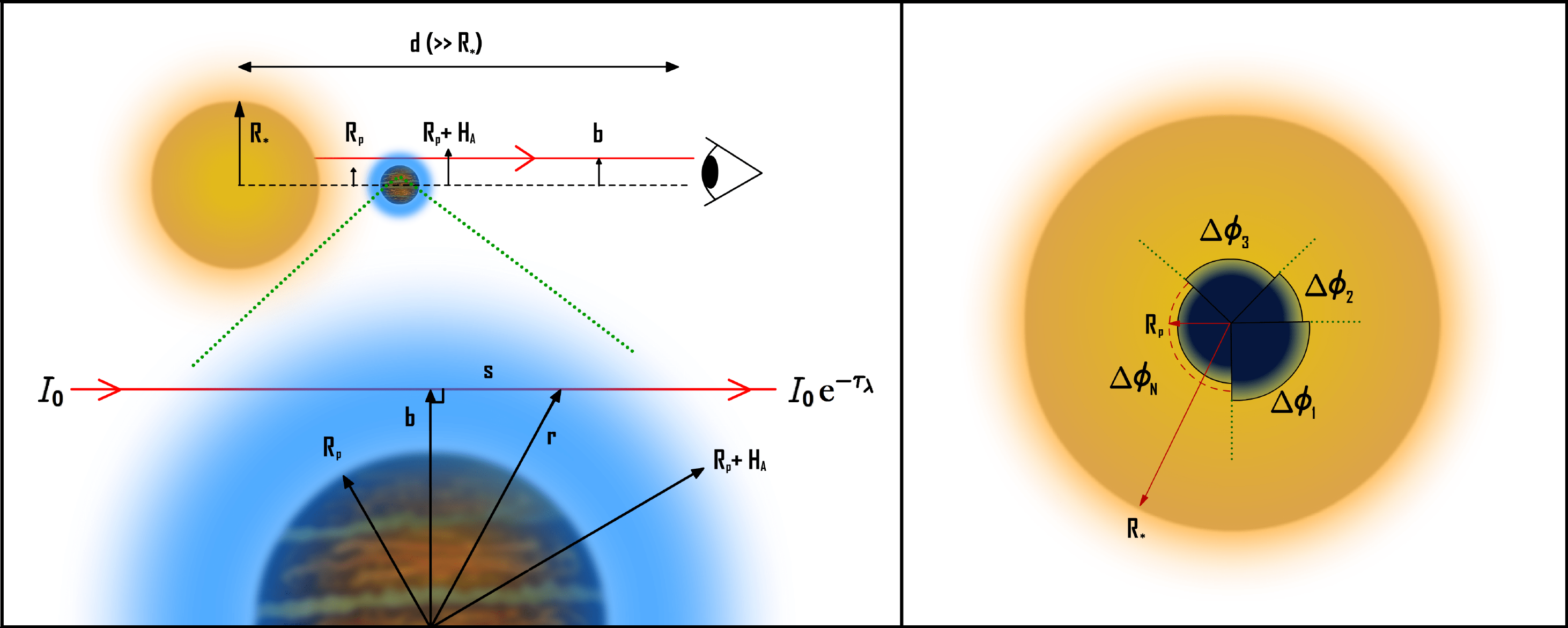}
    \caption{The geometry of a transiting exoplanet. \textbf{Left}: rays of stellar light of an initial intensity $I_0$ at impact parameter $b$ are attenuated due to passage through an atmosphere of height $H_A$. \textbf{Right}: face on projection, with the limb divided into $N$ regions of polar angular extent $\Delta \phi_{n}$, each of which may possess different cloud opacity as a function of $b$ (radial extent exaggerated). Note that the observed transit radius $R_{p}$ represents an average radius at which the atmosphere becomes opaque.}
    \label{fig:Transit}
\end{figure*}

\newpage

\subsection{Forward Model}\label{subsec:forward_model}

Our forward model computes the transmission spectrum of an exoplanet as it transits its host star. We model the day-night terminator of the atmosphere assuming hydrostatic equilibrium and terminator-averaged temperature structure and chemistry. The model allows for inhomogeneous azimuthal cloud and haze distributions. Line-by-line radiative transfer is evaluated under the geometry depicted in Figure \ref{fig:Transit}.

\subsubsection{Radiative Transfer}\label{subsubsec:radiative_transfer}

The transmission spectrum of a generalised two-dimensional atmosphere (derived in Appendix \ref{Appendix:derivation}) is represented by the wavelength-dependant transit depth, $\Delta_{\lambda}$, given by
\begin{equation}
\Delta_{\lambda} = \frac{1}{2 \pi} \int_{0}^{2 \pi} \delta_{\lambda}\, (\phi) ~ d \phi \approx  \sum_{n=1}^{N} \bar{\phi}_{n} \, \delta_{\lambda, n}
	\label{eq:transit_depth_two_dim}
\end{equation}
where $\delta_{\lambda} \, (\phi)$ is the transmission spectrum of an axially symmetric atmosphere with the same composition, temperature structure and cloud properties of the atmosphere at polar angle $\phi$ - i.e. the `one-dimensional' transit depth. We approximate this general case by discretising the atmosphere into $N$ sectors with different properties, specified by the reduced polar angular extent of the $n$th sector, $\bar{\phi}_{n} \equiv \Delta \phi_{n}/2\pi$, (such that $\sum_{n=1}^{N} \bar{\phi}_{n} \equiv 1$). Note that two sectors with different $\bar{\phi}$ but similar properties can be grouped into one. The equivalent 1D transit depth of a sector, $\delta_{\lambda, n}$, is given by
\begin{equation}
\delta_{\lambda, n} = \frac{R_{p}^2 + 2 \displaystyle\int_{R_p}^{R_p+ H_A} b \left(1 - e^{-\tau_{\lambda, n}(b)} \right) db - 2 \displaystyle\int_{0}^{R_p} b e^{-\tau_{\lambda, n}(b)} db}{R_{*}^2}
	\label{eq:transit_depth_one_dim}
\end{equation}
Here, $R_p$ and $R_*$ are the observed radii of the planet and star respectively, $H_A$ is the maximal height of the atmosphere considered (corresponding to $10^{-6}$ bar), $b$ is the impact parameter and $\tau_{\lambda}(b)$ is the optical depth encountered by a ray at a given impact parameter. Equation \ref{eq:transit_depth_one_dim} can be intuitively understood: the first term is achromatic absorption due to an opaque disk of radius $R_p$, the second is absorption due to the annulus of the atmosphere and the third is a correction term accounting for rays that have sufficiently small $\tau_{\lambda}$ to transmit thorough the atmosphere below the observed planetary radius. Though the correction term for rays with $r<R_p$ is often neglected, it must be considered for transmission spectra with two-dimensional clouds -- the observed transit radius represents the average radius at which the planet becomes opaque, which can differ considerably from the local opaque radius in sectors with low opacity (see Figure \ref{fig:Transit}). Note that Equation \ref{eq:transit_depth_two_dim} reduces to Equation \ref{eq:transit_depth_one_dim} in the case $N=1$. In this work we consider only models with $N \leq 2$, leaving the more general case reserved for future work.

To evaluate Equation \ref{eq:transit_depth_one_dim}, we require the optical depth
\begin{equation}
	\tau_{\lambda}(b) = 2\int_{s=0}^{s_{\mathrm{end}}} \kappa_{\lambda}(s') ds' = 2\int_{b}^{R_p + H_A} \kappa_{\lambda}(r) \left(\frac{r}{\sqrt{r^2 - b^2}}\right) dr
\label{eq:tau}
\end{equation}
where $\kappa$ is the \emph{extinction} coefficient, $s$ is the ray path length through the atmosphere and we have suppressed the sector subscript $n$. The extinction is in turn a function of the chemistry and cloud properties in each layer
\begin{equation}
\begin{split}
	\kappa_{\lambda}(r) = & \kappa_{\mathrm{chem}}\, (r) + \kappa_{\mathrm{cia}}\, (r) + \kappa_{\mathrm{cloud}}\, (r) \\
	& \kappa_{\mathrm{\mathrm{chem}}}\, (r) = \sum_{m=1}^{N_{\mathrm{spec}}} n_{m}(r) \, \sigma_{m, \lambda}(r) \\
    & \kappa_{\mathrm{\mathrm{cia}}}\, (r) = \sum_{m=1}^{N_{\mathrm{spec}}} \sum_{l \geq m}^{N_{\mathrm{spec}}} n_{m}(r) \, n_{l}(r) \, \tilde{\alpha}_{ml, \lambda}(r) 
\label{eq:extinction}
\end{split}
\end{equation}
where $n_m$ and $\sigma_{m, \lambda}$ are the number density and absorption cross section of the $m_{th}$ species, $\tilde{\alpha}_{ml, \lambda}$ is the binary absorption coefficient due to collisionally-induced absorption between species $m$ and $l$ and $\kappa_{\mathrm{cloud}}$ is the extinction due to clouds and hazes (defined in section \ref{subsubsec:clouds}). Each of these quantities are written as functions of the radial coordinate $r$, which is short-hand for the pressure and temperature in each layer at which the cross sections are computed. In order to evaluate the total extinction, we thus must specify the pressure-temperature (P-T) profile of the atmosphere.

\subsubsection{Pressure-Temperature Profile}\label{subsubsec:pt_profile}

We divide our model atmosphere into 100 layers spaced uniformly in log-pressure between $10^{-6}$ bar and $10^{\, 2}$ bar. The temperature in each layer is computed via the parametric P-T profile equations given in \citet{Madhusudhan2009}
\begin{align}
\begin{split}
& P = P_{0} \, e^{\, \alpha_1 \sqrt{T-T_0}} \hspace{35pt} (P_0<P<P_1) \\
& P = P_{2} \, e^{\, \alpha_2 \sqrt{T-T_2}} \hspace{35pt} (P_1<P<P_3) \\
& T = T_3 \hspace{94pt} (P>P_3) 
\label{eqn:P-T}
\end{split}
\end{align}
where $P_{0}$ and $T_{0}$ are the pressure and temperature at the top of the atmosphere, $P_{1,3}$ and $T_{1,3}$ are specified at layer boundaries and $P_{2}$ and $T_{2}$ specify the conditions at the (potential) temperature inversion point. The temperature slopes are controlled by $\alpha_1$ and $\alpha_2$. This profile is generic and can include thermal inversions. However, in transmission there is no source function (i.e. thermal emission), which means inversions are unphysical. We account for this taking $P_{2} \leq P_{1}$.

Once $P$ and $T$ are specified in each layer, the total number density, $n_{\mathrm{tot}}(r)$, and $r$ are determined by the ideal gas law and hydrostatic equilibrium. This requires a reference pressure, $P_{\mathrm{ref}}$, to be specified, which we take to be the pressure at which $r=R_p$. The gravitational field follows an inverse square law over radial distance with $g(R_p) = 9.192 \, \mathrm{m \, s^{-1}}$ and $R_p = 1.359 R_J$ for HD 209458b. We further assume a $\mathrm{H}_{2}-\mathrm{He}$ dominated atmosphere with mean molecular weight per particle $\mu=2.3$. Taking into account continuity at boundaries, and considering only average terminator profiles, our parametric P-T profile is specified by 7 parameters: ($\alpha_1, \, \alpha_2, \, T_0, \, P_1, \, P_2, \, P_3, \, P_{\mathrm{ref}}$).

\subsubsection{Chemistry}\label{subsubsec:chemistry}

We consider the main chemical species regarded as spectrally active between $0.3-1.7 \micron$ in hot Jupiter atmospheres: $\mathrm{H_2}, \,  \mathrm{He}, \, \mathrm{H_{2}O}, \,  \mathrm{CH_{4}}, \,  \mathrm{NH_{3}}, \,  \mathrm{HCN}, \,  \mathrm{CO}, \, \mathrm{CO_{2}}, \,  \mathrm{Na}$ and $\mathrm{K}$ \citep{seager2000,Madhusudhan2016a}. We quantify the abundance of each chemical species by its mixing ratio: $X_{i} \equiv n_{i}/n_{\mathrm{tot}}$, which we assume to be uniform in the observable atmosphere. We ascribe a single mixing ratio for each species in the day-night terminator region of the atmosphere -- effectively corresponding to the average limb abundance. 

\begin{figure}
	\includegraphics[width=\columnwidth,  trim={0.6cm 0.8cm 0.4cm 0.6cm}]{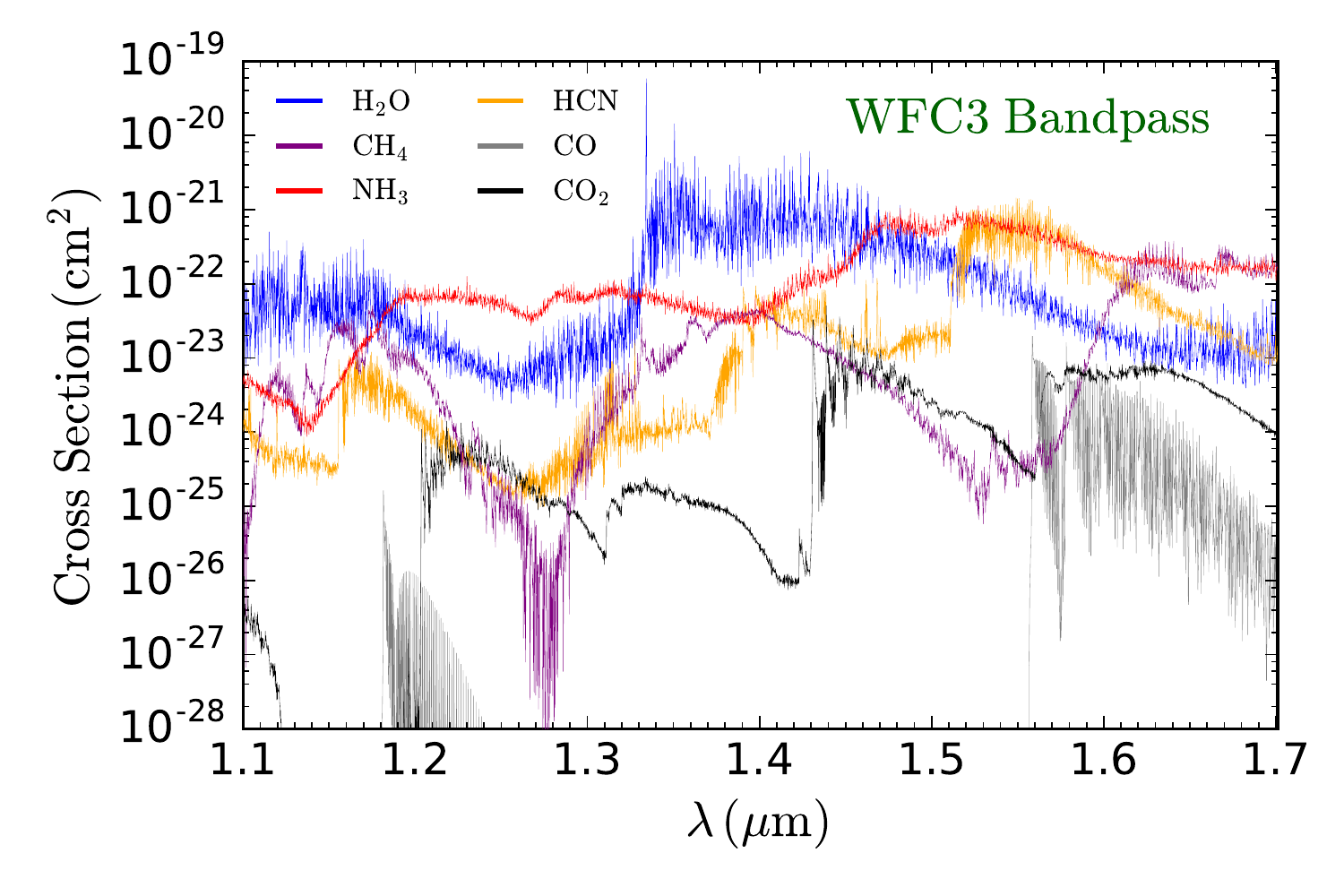}
    \caption{Cross sections of spectrally active molecules in the HST WFC3 bandpass, evaluated at 1400 K and $10^{-3}$ bar -- representative of the upper atmosphere of HD 209458b probed in transmission.}
    \label{fig:cross_sections}
\end{figure}

The molecular cross sections are pre-computed line-by-line following the methodology outlined in \citet{Hedges2016} and Gandhi \& Madhusudhan (submitted). We use the latest theoretical and experimental line lists available, drawing from the HITEMP database for $\mathrm{H_{2}O}, \,  \mathrm{CO}$ and $\mathrm{CO_{2}}$ \citep{Rothman2010} and EXOMOL for $\mathrm{CH_{4}}$, HCN and $\mathrm{NH_{3}}$ \citep{Tennyson2016}. The pre-computed cross sections are binned down to a resolution of 1 $\mathrm{cm^{-1}}$ on a pre-defined temperature and pressure grid ranging from $10^{-4} - 10^{\, 2}$ bar and $300 - 3500$ K. Given a P-T profile, POSEIDON interpolates these cross sections linearly in log-pressure and temperature to those in each atmospheric layer. Figure \ref{fig:cross_sections} shows the resulting molecular cross sections at a representative temperature and pressure of $1400$ K and $10^{-3}$ bar. For $\mathrm{Na}$ and $\mathrm{K}$, we use the cross sections used in \citet{Christiansen2010}, based on semi-analytic Lorentzian line profiles, which will be replaced with more accurate cross sections in future work. For collisionally-induced absorption, we consider $\mathrm{H}_{2} - \mathrm{H}_{2}$ and $\mathrm{H}_{2} - \mathrm{He}$ absorption from the HITRAN database \citep{Richard2012a}. 

\subsubsection{A Generalised 2D Cloud/Haze Prescription}\label{subsubsec:clouds}

\begin{figure*}
	\includegraphics[width=\linewidth,  trim={0.5cm 0.6cm 0.2cm 0.4cm}]{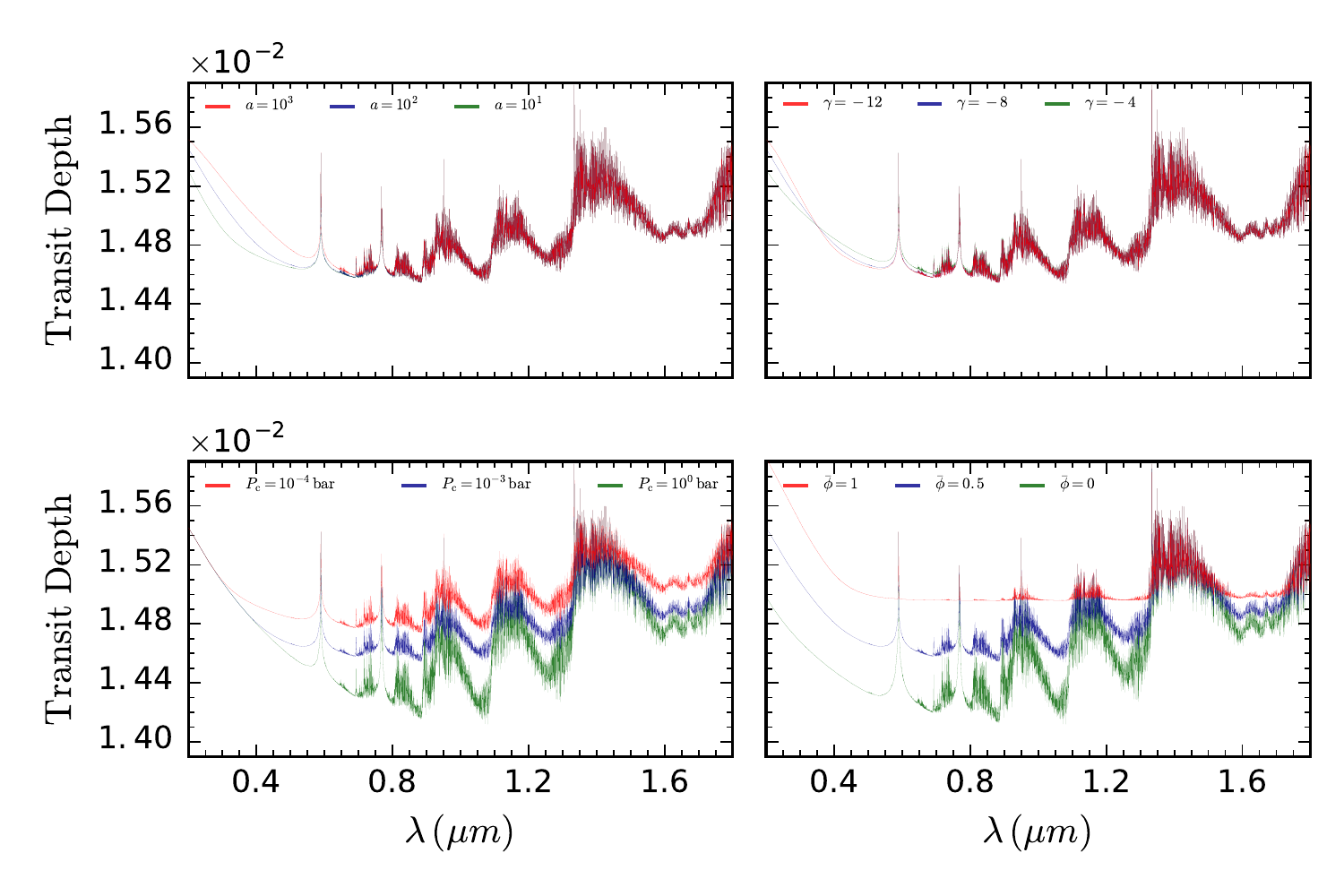}
    \caption{The variation of an $R \approx 10000$ transmission spectrum (blue) with cloud/haze properties. The top panels show progressively enhanced scattering in the optical due to hazes, whilst the bottom panels show the variation with the opaque cloud deck pressure (left) and terminator cloud fraction (right). In all cases, the atmosphere is solar-composition at an isothermal temperature of 1400 K.}
    \label{fig:cloud_parameters}
\end{figure*}

Correctly modelling exoplanet transmission spectra requires the inclusion of clouds and hazes. We consider `cloudy' regions of our atmosphere to consist of an opaque cloud deck at pressures $P \geq P_{\mathrm{cloud}}$ and scattering due to hazes above the clouds  \citep{Etangs2008}. The extinction is given by

\begin{equation}
\kappa_{\mathrm{\mathrm{cloud}}} \, (r)=
\left\{
\begin{array}{ll}
    a \, \sigma_{0} (\lambda/\lambda_{0})^{\gamma}  \hspace{35pt} (P < P_{\mathrm{cloud}}) \\
    \infty        \hspace{72pt} (P \geq P_{\mathrm{cloud}}) 
\end{array}
\right.
\label{eq:cloud_extinction}
\end{equation}
where $\lambda_0$ is a reference wavelength (here, 350 nm), $\sigma_0$ is the $\mathrm{H}_2$-Rayleigh scattering cross section at the reference wavelength ($5.31 \times 10^{-31} \, \mathrm{{m}^2}$), $a$ is the `Rayleigh-enhancement factor' and $\gamma$ is the `scattering slope'. The first term in Equation \ref{eq:cloud_extinction} accounts for scattering due to high-altitude hazes, whilst the second term models the sharp cut-off in transmission due to the high optical depths encountered inside clouds in the transit geometry \citep{Fortney2005}.

POSEIDON accounts for the possibility of generalised azimuthally-dependant terminator cloud coverage (Figure \ref{fig:Transit}). In this work, we take $N=2$ in Equation \ref{eq:transit_depth_two_dim} and consider region 1 to contain clouds/hazes according to Equation \ref{eq:cloud_extinction} and region 2 to be clear. In this limit, we recover the `patchy cloud' case given in \citet{Line2016a}
\begin{equation}
\Delta_{\lambda} = \bar{\phi} \, \delta_{\lambda, \mathrm{cloudy}} + (1 - \bar{\phi}) \, \delta_{\lambda, \mathrm{clear}}
	\label{eq:transit_depth_patchy}
\end{equation}
where we have defined $\bar{\phi}_1 \equiv \bar{\phi}$, $\delta_{\lambda, 1} \equiv \delta_{\lambda, \mathrm{cloudy}}$, $\delta_{\lambda, 2} \equiv \delta_{\lambda, \mathrm{clear}}$ and used $\bar{\phi}_2 =  1 - \bar{\phi}_1$ to make the correspondence clear. The reduced polar angle $\bar{\phi}$ then encodes the total terminator cloud coverage (though the cloud need not be distributed continuously with a sharp boundary at the clear interface). In clear regions, we arbitrarily set $P_{\mathrm{cloud}} = 50 \, \mathrm{bar}$ (in order to ensure it has no effect on the transmission spectrum) and consider scattering in the visible to be solely due to molecular $\mathrm{H}_2$ - using the cross section given in \citet{Dalgarno1962}.

Figure \ref{fig:cloud_parameters} shows the effect of varying our cloud-parameters ($a, \, \gamma, \, P_{\mathrm{cloud}}, \, \bar{\phi}$) on a default fiducial transmission spectrum (blue) generated by the POSEIDON forward model. The top panels demonstrate that $a$ and $\gamma$ encode the strength and slope, respectively, due to scattering, which tends to manifest most prominently at visible wavelengths $\lambda < 0.7 \micron$. In contrast, the altitude of the opaque cloud and the cloud coverage along the terminator strongly influence spectra across both the optical and infrared. Raising the cloud deck (lowering $P_{\mathrm{cloud}}$) results in an increased transit depth at all wavelengths, as the $\tau=1$ surface is pushed to progressively higher altitudes. As $\bar{\phi} \rightarrow 1$, the base-level of the spectrum becomes increasingly flat until the one-dimensional limit of a uniform cloud acting to shrink the transmitting annulus of the atmosphere is recovered.

With the combination of temperature structure, chemistry and cloud properties now all specified, Equation \ref{eq:transit_depth_two_dim} can be computed and the transmission spectra evaluated. By specifying this model in terms of parameters, a large ensemble of spectra can be evaluated for different atmospheric properties. We now turn to the essence of retrieval: the usage of a statical algorithm to extract atmospheric properties.

\subsection{Statistical Retrieval}\label{subsec:stats}

\begin{figure}
	\includegraphics[width=\columnwidth,  trim={0.5cm 0.6cm 0.8cm 0.8cm}]{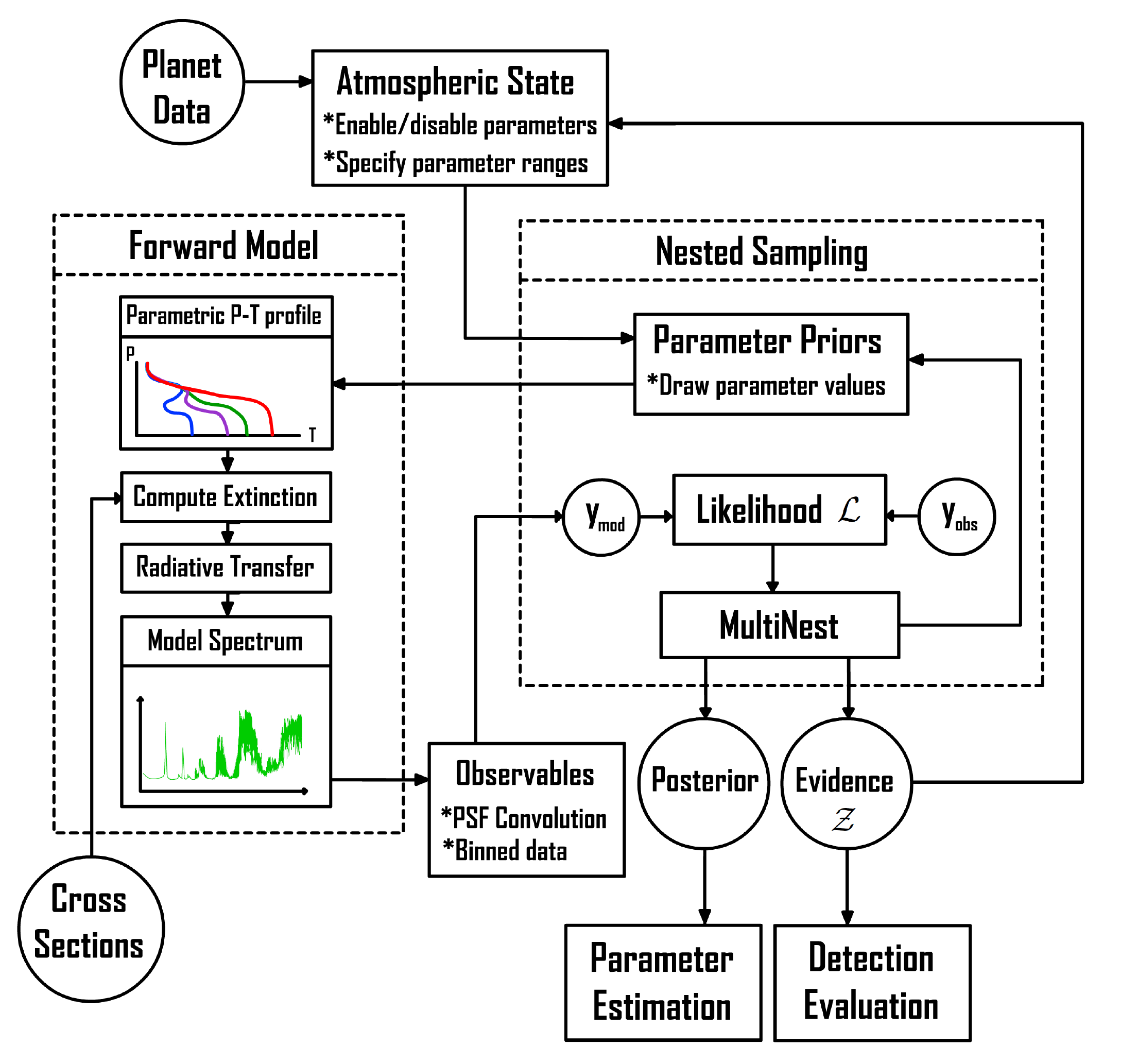}
    \caption{Architecture of the atmospheric retrieval algorithm POSEIDON. A forward model is repeatedly called to generate transmission specta for different parameter inputs -- each selected and guided by a nested sampling algorithm. The output is a set of posterior samples and the Bayesian evidence.}
    \label{fig:Architecture}
\end{figure}

Ultimately, we are interested in extracting the underlying properties of an exoplanet atmosphere -- i.e, the values of the parameters underlying the forward model (P-T profile, chemistry, clouds etc.) -- from observed transmission spectra. An additional question one must assess is the suitability of the forward model itself in light of the data. These two tasks, parameter estimation and model comparison, can be accomplished within a Bayesian framework.

The basic architecture of our retrieval algorithm is depicted in Figure \ref{fig:Architecture}. The atmosphere is encoded by a vector of underlying physical parameters. For a given parameter combination, the forward model outputs a spectrum that is convolved with relevant instrument point spread functions (PSF) and/or integrated over the respective instrument functions to produce predicted data points. At each point in parameter space, these predicted data points, denoted by $y_{\mathrm{mod}}$, are compared with the observed data points, $y_{\mathrm{obs}}$, to compute the likelihood of the given set of parameters. The likelihood in turn informs the choice of the next set of parameters by the statistical retrieval algorithm, depicted on the right of Figure \ref{fig:Architecture}. The algorithm allows thorough exploration of the entire multi-dimensional parameter space and computation of the Bayesian evidence -- which quantifies the suitability of the model itself. POSEIDON employs a nested sampling algorithm to accomplish this purpose.

We proceed to summarise the parametrisation of our forward model in section \ref{subsubsec:parameters}, define the statistical aspects and terminology of atmospheric retrieval in section \ref{subsubsec:Bayesian_stats} and, finally, describe the nested sampling statistical algorithm employed by POSEIDON in section \ref{subsubsec:nested_sampling}.

\subsubsection{Atmospheric Parametrisation}\label{subsubsec:parameters}

In this implementation, the POSEIDON forward model is described by up to 16 parameters: 7 for the terminator P-T profile, 5 for the terminator chemistry ($X_{\mathrm{H_{2}O}} \, , \,  X_{\mathrm{CH_{4}}} \, , \, X_{\mathrm{NH_{3}}} \, , \,  X_{\mathrm{HCN}}  \, , \, X_{\mathrm{Na}}$) and a 4-parameter cloud/haze description. The allowed range and Bayesian prior used for each parameter are given in Table \ref{table:priors}. Note that we have restricted our choice of molecules to those expected to dominate absorption in the WFC3 bandpass (see Figure \ref{fig:cross_sections}), thus the retrieval results we present do not include $ X_{\mathrm{CO}}$ and $X_{\mathrm{CO_{2}}}$. We have further elected not to describe the K abundance by an additional parameter, given the simplicity of our current alkali cross section implementation, instead fixing the K/Na ratio to the solar value ($\approx 6.4 \times 10^{-2}$). Since we focus on hot Jupiters, we assume a $\mathrm{H}_{2}-\mathrm{He}$ dominated atmosphere with a fixed $\mathrm{He}/\mathrm{H}_{2}$ ratio of 0.17.

\newcommand{\ra}[1]{\renewcommand{\arraystretch}{#1}}
\begin{table}
\ra{1.3}
\caption{Retrieval parameters and prior probabilities}
\begin{tabular*}{\columnwidth}{l@{\extracolsep{\fill}} llll@{}}\toprule
$\mathrm{Parameter}$ & $\mathrm{Prior}$ & $\mathrm{Range}$\\ \midrule
$\alpha_{1,2}$ & Uniform & $0.02 \ \mathrm{to} \ 1 \ \mathrm{K}^{-1/2}$  \\
$T_{0}$ & Uniform & $800 \ \mathrm{to} \ 1600 \ \mathrm{K}$ \\
$P_{1,2}$ & Log-uniform & $10^{-6} \ \mathrm{to} \ 10^{\,2} \ \mathrm{bar}$ \\
$P_{3}$ & Log-uniform & $10^{-2} \ \mathrm{to} \ 10^{\,2} \ \mathrm{bar}$ \\
$P_{\mathrm{ref}}$ & Log-uniform & $10^{-4} \ \mathrm{to} \ 10^{\,2} \ \mathrm{bar}$ \\
$X_{i}$ & Log-uniform & $10^{-10} \ \mathrm{to} \ 10^{-2}$ \\
$a$ & Log-uniform & $10^{-4} \ \mathrm{to} \ 10^{\,8}$ \\
$\gamma$ & Uniform & $-20 \ \mathrm{to} \ 2$ \\
$P_{\mathrm{cloud}}$ & Log-uniform & $10^{-6} \ \mathrm{to} \ 10^{\,2} \ \mathrm{bar}$ \\
$\bar{\phi}$ & $\text{Uniform}^{\dagger}$ & $0 \ \mathrm{to} \ 1$ \\
\bottomrule
\vspace{0.1pt}
\end{tabular*}
$^{\dagger} \, $ For additional sectors, a Dirichlet prior is more appropriate.
\label{table:priors}
\end{table}

We elect for generous `uninformative' priors. A uniform prior probability distribution (section \ref{subsubsec:Bayesian_stats}) is ascribed to parameters expected to vary by less than two orders of magnitude, whilst a uniform-in-the-logarithm prior is used for parameters expected to vary over many orders of magnitude. We elect in this initial work to restrict ourselves to $N=2$ atmospheric sectors, in which case only 1 parameter, $\bar{\phi}$, is required to describe two-dimensional effects. An additional subtlety in the choice of these priors must be made explicit: since both the mixing ratios $X_{i}$ and the reduced polar extent $\bar{\phi_{i}}$ are subject to the constraint of addition to unity, in the most general case a flat Dirchlet prior (uniform over a simplex subspace) is most appropriate. Here we use the fact that we know a priori that the dominant component of hot Jupiter atmospheres is $\mathrm{H}_{2}-\mathrm{He}$ to treat the remaining gases as trace species with log-uniform priors -- this assumption must be relaxed for high mean molecular weight atmospheres, such as those of super-Earths \citep{Benneke2012}. Similarly, for a two-sector atmosphere we need only ascribe a single uniform parameter $\bar{\phi}_{1}$, with the extent of the second sector automatically specified by the unity summation condition.

\subsubsection{Bayesian Framework}\label{subsubsec:Bayesian_stats}

Consider a set of forward models $M_{i}$ described by an underlying set of physical parameters $\boldsymbol{\theta}$. Our a priori expectations on the values of the parameters are encoded in the \emph{prior probability density function}: $p \left(\boldsymbol{\theta} \middle| M_{i} \right)$. By obtaining a set of observations $\mathbfit{y}_{\mathrm{obs}}$, we can formally update our knowledge on the values of these parameters via Bayes' theorem
\begin{equation}
p \left(\boldsymbol{\theta} \middle| \mathbfit{y}_{\mathrm{obs}} \, , M_{i} \right) = \frac{p \left(\mathbfit{y}_{\mathrm{obs}} \middle| \boldsymbol{\theta} \, , M_{i} \right) \, p \left(\boldsymbol{\theta} \middle| M_{i} \right)}{p \left(\mathbfit{y}_{\mathrm{obs}} \middle| M_{i} \right)} \equiv \frac{\mathcal{L}\left(\mathbfit{y}_{\mathrm{obs}} \middle| \boldsymbol{\theta} \, , M_{i} \right) \, \pi\left(\boldsymbol{\theta} \middle| M_{i} \right)}{\mathcal{Z}\left(\mathbfit{y}_{\mathrm{obs}} \middle| M_{i} \right)}
	\label{eq:bayes_theorem}
\end{equation}
where in the equivalency we have defined the conventional notation for the \emph{likelihood function}, $\mathcal{L}$, prior, $\pi$, and \emph{Bayesian evidence}, $\mathcal{Z}$. The quantity on the left-hand side is the \emph{posterior probability density function}, which quantifies our knowledge on the values of the parameters of model $M_{i}$ following an observation. The priors for each parameter are given in Table \ref{table:priors}, whilst the likelihood -- a measure of the plausibility of the forward model producing the observed data for a choice of model parameters -- is given, for observations with independently distributed Gaussian errors, by
\begin{equation}
\mathcal{L}\left(\mathbfit{y}_{\mathrm{obs}} \middle| \boldsymbol{\theta} \, , M_{i} \right) = \prod_{k=1}^{N_{\mathrm{obs}}} \frac{1}{\sqrt{2\pi}\sigma_{k}} \mathrm{exp}\left(-\frac{[y_{\mathrm{obs} \,, \,k} - y_{\mathrm{mod} \,, \,k} \,(\boldsymbol{\theta})]^2}{2\sigma_{k}^2}\right)
	\label{eq:likelihood}
\end{equation}
where $\mathbfit{y}_{\mathrm{mod}} \,(\boldsymbol{\theta})$ are the set of (binned) model data points produced by convolving the output of the forward model with the relevant instrument PSFs and integrating over the corresponding instrument functions.

The Bayesian evidence is the key quantity employed in Bayesian model comparison. From Equation \ref{eq:bayes_theorem}, we see that it is simply the normalising factor that ensures the integral of the posterior probability density over the entire parameter space evaluates to unity
\begin{equation}
\mathcal{Z}\left(\mathbfit{y}_{\mathrm{obs}} \middle| M_{i} \right) = \int_{\mathrm{all} \, \boldsymbol{\theta}} \mathcal{L}\left(\mathbfit{y}_{\mathrm{obs}} \middle| \boldsymbol{\theta} \, , M_{i} \right) \pi\left(\boldsymbol{\theta} \middle| M_{i} \right) \, \mathrm{d}\boldsymbol{\theta}
	\label{eq:evidence}
\end{equation}
The value of the evidence can be qualitatively understood as a `figure of merit' that is maximised by models with a high likelihood in a compact parameter space \citep{Trotta2008}. To see this, consider the addition of a new parameter to a model. By extending the dimensionality of the parameter space, the value of the prior probability density $\pi$ will be diluted across this additional volume. The evidence for this more complex model will then only increase if the new volume contains previously unsampled regions of high likelihood. In contrast to simply fitting a spectrum, the Bayesian evidence provides an automatic implementation of Occam's Razor by penalising models with unjustified complexity.

The quantitative utility of the evidence becomes clear when using Bayes' theorem to consider the relative probability of two competing models in light of the data
\begin{equation}
\frac{p \left(M_{i} \middle| \mathbfit{y}_{\mathrm{obs}} \right)}{p \left(M_{j} \middle| \mathbfit{y}_{\mathrm{obs}}\right)} = \frac{\mathcal{Z}\left(\mathbfit{y}_{\mathrm{obs}} \middle| M_{i} \right)}{\mathcal{Z}\left(\mathbfit{y}_{\mathrm{obs}} \middle| M_{j} \right)} \, \frac{p \left(M_{i} \right)}{p \left(M_{j} \right)} \equiv \mathcal{B}_{ij} \frac{p \left(M_{i} \right)}{p \left(M_{j} \right)}
	\label{eq:posterior_odds}
\end{equation}
where in the equivalency we have defined the \emph{Bayes factor} for model $i$ vs. model $j$. We assume the final factor, expressing their a priori odds ratio, to be unity. Thus the ratio of the evidences between two models allows the adequacy of the models themselves to be assessed. Values of at least $\mathcal{B}_{ij} = 3, 12, 150$ are often interpreted as `weak', `moderate' and `strong' detections in favour of $M_{i}$ over $M_{j}$ \citep{Trotta2008,Benneke2013}. The Bayes factor can in turn be related to the commonly used frequentist measure of sigma-significance \citep{Selke2001}, which we also communicate when presenting Bayes factors during model comparison.

Once an adequate  model is identified via Bayesian model comparison, one can then constrain the parameters within the chosen model. This is accomplished by drawing samples from the posterior (Equation \ref{eq:bayes_theorem}) and marginalising (integrating) over the full range of the other parameters. The resulting probability density histograms for each parameter encode our knowledge of the underlying physics of the atmosphere. Strictly speaking, the Bayesian evidence is not required if parameter estimation is the only goal (as $\mathcal{Z}$ only normalises the posterior), though the implicit assumption in this case is that the model itself is `correct'.

\subsubsection{Nested Sampling}\label{subsubsec:nested_sampling}

The statistical algorithms employed by retrieval codes are becoming increasingly sophisticated. Early retrievals used grid-based parameter space exploration \citep{Madhusudhan2009} or optimal estimation techniques \citep{Lee2012,Line2012,Barstow2013} that do not allow for full marginalisation of the posterior. Rigorous parameter estimation in a Bayesian framework was enabled by the integration of Markov Chain Monte Carlo (MCMC) algorithms into retrieval codes \citep[e.g.][]{Madhusudhan2011,Benneke2012,Line2013}. However, these techniques do not permit a computationally efficient evaluation of the multi-dimensional integral in Equation \ref{eq:evidence}, and so renders Bayesian model comparison challenging. Atmospheric retrieval is currently undergoing a phase-transition, with the technique of nested sampling \citep{Skilling2004} -- which allows efficient computation of both the Bayesian evidence in addition to providing posterior samples for parameter estimation -- now utilised in the majority of applications \citep{Benneke2013,Benneke2015,Waldmann2015a,Line2016a,Lupu2016a,Lavie2016}.

POSEIDON implements the multimodal nested sampling algorithm MultiNest \citep{Feroz2008,Feroz2009,Feroz2013} via the Python wrapper PyMultiNest \citep{Buchner2014}. MultiNest computes the Bayesian evidence numerically by transforming Equation \ref{eq:evidence} to a one-dimensional integral; evaluated by sweeping through progressively increasing iso-likelihood contours containing a set of `live points' drawn from progressively shrinking ellipsoids. The full history of these samples serves to widely chart the posterior -- see \citet{Benneke2013} for a detailed discussion of the algorithm. A notable strength of MultiNest in the context of atmospheric retrieval is its ability to navigate significantly non-Gaussian, degenerate and non-trivially curved posteriors, in addition to being fully parallelised for cluster computing.

By coupling the POSEIDON forward model to MultiNest, we are able to assess the plausibility of a wide variety of physical phenomena (e.g. detections of chemical species, hazes, clear vs. cloud-free atmospheres). For each model, we derive the posterior probability distributions of the underlying parameters, which typically requires the evaluation of many millions of spectra to obtain robust parameter estimates. Now that our retrieval framework has been established, we proceed to demonstrate its effectiveness.

\section{Validation}\label{section:validation}

 There are two key steps to validate a retrieval code: i) verify that the forward model outputs correct spectra; ii) successfully retrieve the atmospheric state underlying a simulated dataset produced by the forward model. Tests of this manner enable the accuracy and reliability of a retrieval code to be established, as well as bringing to light any potential biases or degeneracies in the results that could impact the interpretation of real spectra.

\begin{figure*}
	\includegraphics[width=\textwidth, trim={0.16cm 0.6cm 0.16cm 0.45cm},clip]{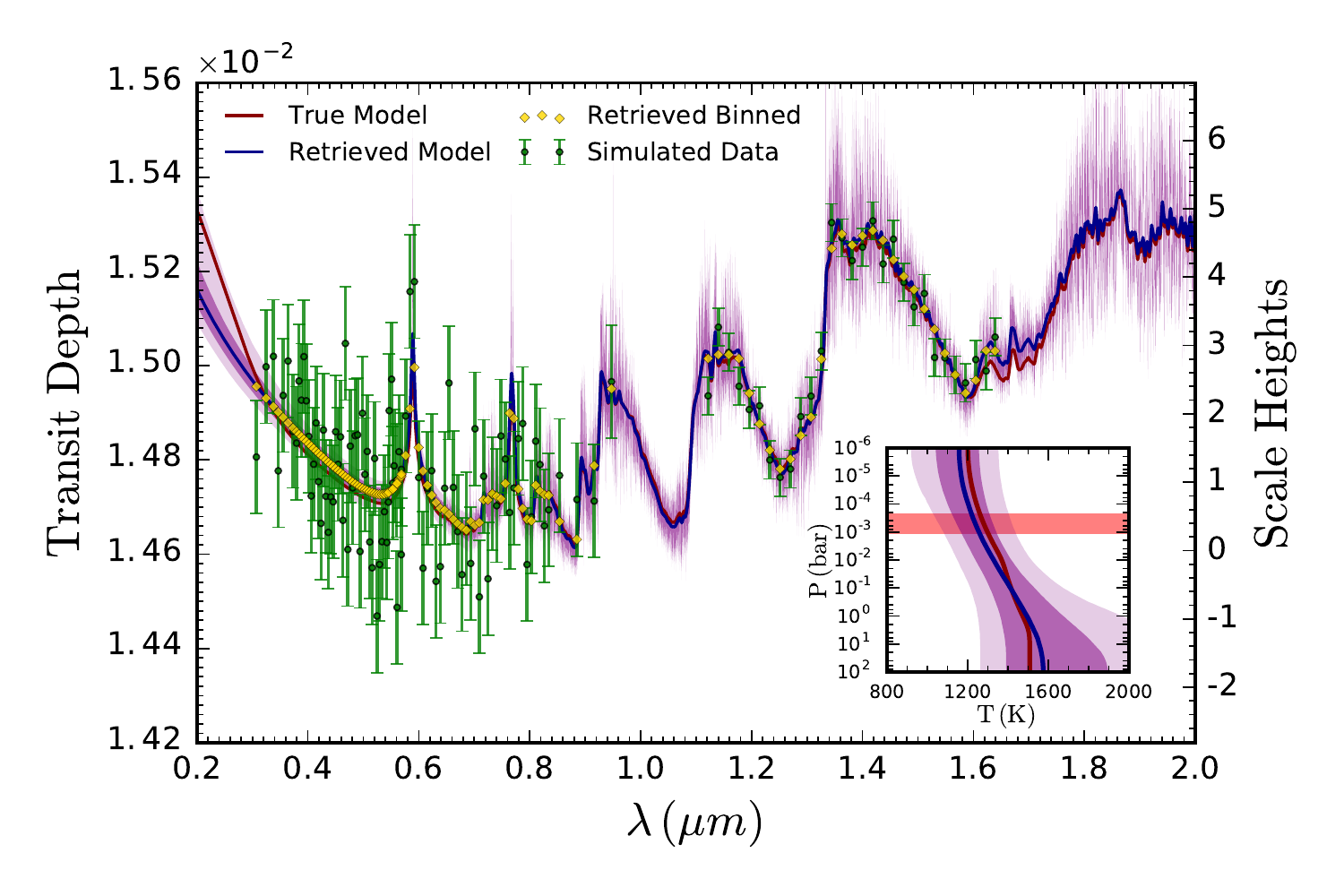}
    \caption{Validation of POSEIDON's ability to retrieve a spectrum and P-T profile from a synthetic data set. $\textbf{Main figure:}$ a high resolution ($R \approx 10000$) spectrum is generated by the forward model, convolved with the WFC3 G141 \& STIS G430 / G750 PSFs and integrated over the corresponding instrument functions to produce binned synthetic model points. Gaussian errors of 120ppm and 40ppm in the optical and near-infrared are added to create synthetic observations, shown in green. The yellow diamonds are the median binned model points resulting from the retrieval. The red and blue curves are Gaussian smoothed representations of the true spectrum and median retrieved spectrum, respectively. The dark and light purple regions indicate $1\sigma$ and $2\sigma$ confidence regions in the transit depth at each wavelength, derived from 10,000 random samples drawn from the posterior distribution. The number of equivalent scale heights above $(R_{p}/R_{*})^2$ is computed with respect to the median retrieved photosphere temperature. $\textbf{Inset:}$ retrieved terminator P-T profile. The red and blue curves are the true and median retrieved temperature profiles, respectively, the purple regions $1\sigma$ and $2\sigma$ confidence regions for the temperature in each layer and the red region the $1\sigma$ extent of the near-infrared photosphere ($\tau = 1$ at $1.5 \micron$).}
    \label{fig:sim_retrieved_spectrum}
\end{figure*}

\subsection{Forward Model Validation}\label{subsection:forward_model_validation}

Prior to the generation of a simulated dataset, we undertook an internal comparison between the POSEIDON forward model and that utilised in \citet{Madhusudhan2014c}. Though the radiative transfer schemes and parametric P-T profiles in both models are identical, to make a direct comparison the cloud and haze parametrisation built into POSEIDON was set to a cloud-free atmosphere (i.e. $\bar{\phi}=0$) and we temporarily replaced our new cross sections with those employed in \citet{Madhusudhan2014c}. We tested our model over a range of parameters, finding only a minor offset ($\sim 10^{-5}$) in the transit depth due to the higher order numerical methods employed by POSEIDON.

\begin{figure*}
	\includegraphics[width=\linewidth, trim={-0.85cm .05cm -1.25cm 0.0cm},clip]{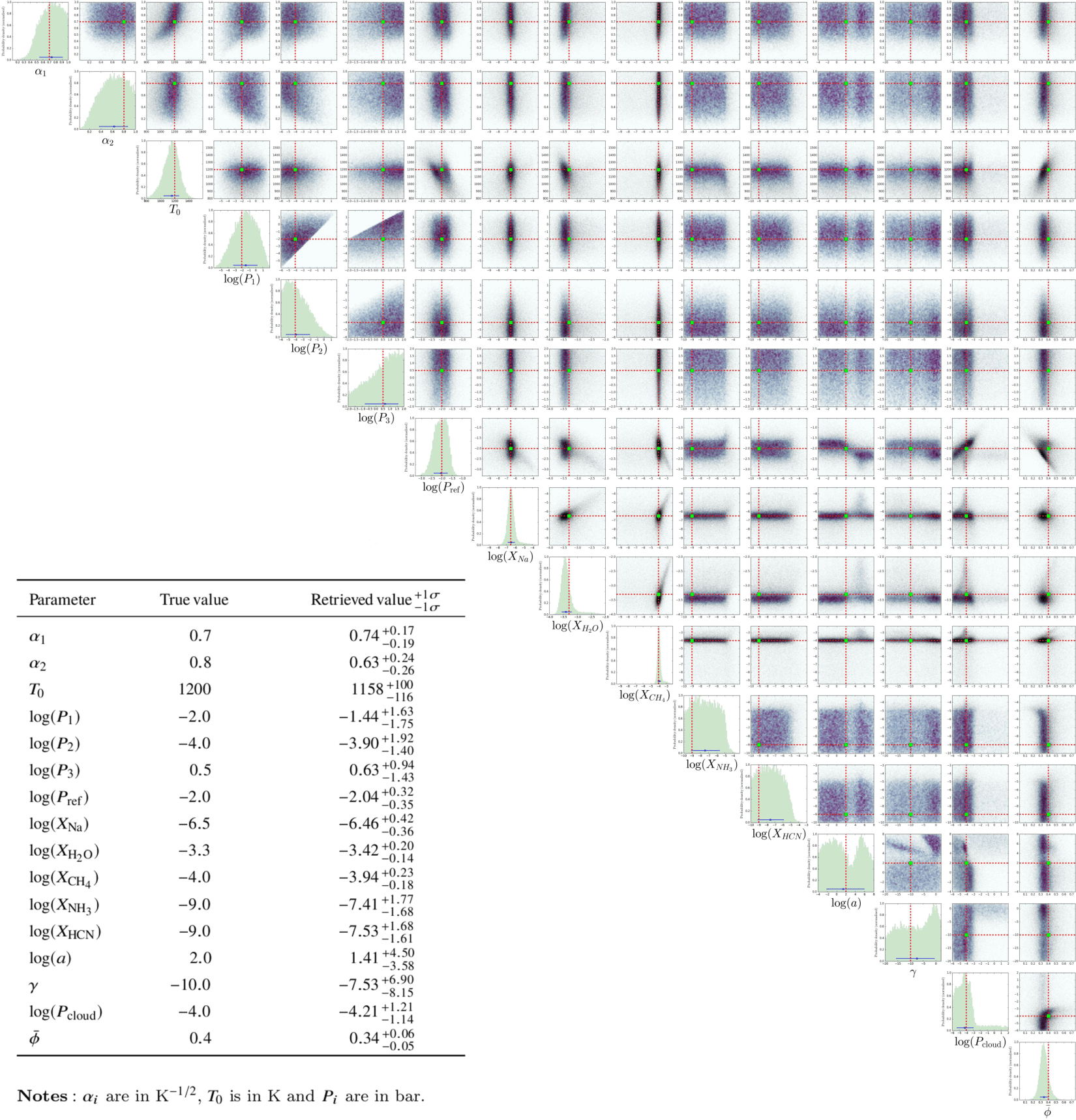}
    \caption{Full posterior distribution from POSEIDON's retrieval of the atmospheric properties underlying the synthetic data in Figure \ref{fig:sim_retrieved_spectrum}. $\textbf{Main figure:}$ corner plot depicting correlations between pairs of retrieved parameters and marginalised histograms for each parameter. The red dashed lines and green squares indicate the true values of each parameter. $\textbf{Table inset:}$ the true parameter values used to generate the simulated dataset and the corresponding median retrieved values. All parameters are correctly retrieved to within $1\sigma$.}
    \label{fig:sim_posterior}
\end{figure*}

\subsection{Retrieval Validation}\label{subsection:retrieval_validation}

We now proceed to demonstrate the typical results from an application of the POSEIDON retrieval code to a given dataset. The goal here is to start with a synthetic data set, based on a known model spectrum, and use POSEIDON to retrieve the underlying model parameters. This allows us to test how well the parameters can be obtained. 

We first generated a high-resolution fiducial solar-composition transmission spectrum, using the planetary properties of HD 209458b. The P-T profile parameters were chosen to produce a monotonically decreasing temperature with altitude, such that the temperature in the photosphere ($P \sim 10^{-2}$ bar) is around that of the planetary equilibrium temperature ($T_{\mathrm{phot}} \approx 1400$ K). We ascribed a cloud coverage of 40 per cent to the terminator, with cloudy regions consisting of a high altitude opaque cloud deck at 0.1 mbar subsumed in a uniform-with-altitude haze. The values of the physical parameters used to produce the spectrum are given in the embedded table in Figure \ref{fig:sim_posterior}. 

With the spectrum generated, we produced a simulated dataset at a precision commensurate with currently available observations. We first convolved the high-resolution spectrum with the relevant PSFs for the HST STIS (G430/G750) and WFC3 (G141) instruments and integrated over the corresponding instrument functions to produce a set of low-resolution binned spectral points at the same wavelength locations as the HD 209458b data given in \citet{Sing2016}. We added Gaussian errors at the levels of 120ppm and 40ppm in the visible and near-infrared respectively to produce simulated STIS and WFC3 data points. The combined dataset served as the input to POSEIDON.

We initialised multiple nested sampling runs with 1,000-8,000 live points in order to confirm consistency. To efficiently compute the millions of models required to adequately explore the entire 16 dimensional parameter space, we evaluated transmission spectra at 2000 wavelengths uniformly spaced over the range 0.2-2.0 $\micron$. We show the full posterior for the retrieval with 4,000 live points in Figure \ref{fig:sim_posterior}. We find that this number of live points offers an optimum trade-off between minimising the error in the computed Bayesian evidence ($\Delta \ln \mathcal{Z} \approx 0.05$) and optimising the overall time necessary to reach the convergence criteria.

The median retrieved spectrum shows excellent agreement with the true spectrum. Figure \ref{fig:sim_retrieved_spectrum} shows that they are coincident throughout the spectral range 0.3-2.0 $\micron$ to < 40ppm precision -- even where there is limited data coverage. The WFC3 bandpass is constrained even more tightly, typically to < 20ppm precision. The only region of significant deviation is in the UV below 0.3 $\micron$, where there are no data points to inform the retrieval.

The true P-T profile agrees with median retrieved profile to < 50 K. Indeed, the $1\sigma$ extent of the profile is tightly constrained to $\sim 100$ K, with the contours matching the overall shape of the profile with altitude. As expected, the constraint becomes tighter around the photosphere ($P \sim 10^{-2}$ bar), where the information content is greatest, and expands at altitudes away from those probed in transmission. This demonstrates that the terminator P-T profile shape can be correctly inferred from high-precision transmission spectra.

Figure \ref{fig:sim_posterior} demonstrates that POSEIDON correctly retrieves all the parameters used to generate our simulated dataset to $1\sigma$. The most tightly constrained parameters are the abundances of $\mathrm{H_{2}O}$ and $\mathrm{CH_{4}}$, which are on the order of 0.2 dex. Where absorption signatures of a molecular species are not deemed necessary to explain the data, such as with the low $\mathrm{NH_{3}}$ and $\mathrm{HCN}$ abundances, the posterior retains the flat shape of the prior below an established upper bound. Though the mode of the cloud deck pressure is coincident with the true value, and the terminator cloud fraction is sharply localised, the haze properties remain relatively unconstrained at the noise level of the optical data. This is unsurprising, as light transmitting through the relatively small fraction of the model atmosphere above the 0.1 mbar cloud deck will be insensitive to the haze. Indeed, the small tails in the chemical abundances to higher values can be seen as stemming from a weak correlation with the possibility of strong ($a \approx 10^{\, 6}$) hazes. For data where the scattering slope can be readily resolved, we do not see this tail.

\subsubsection{Breaking Cloud-Composition Degeneracies}\label{subsubsection:cloud_degeneracy}

Chemical inferences are relatively independent of the values of the cloud parameters. This can be seen in Figure \ref{fig:sim_posterior} by the roughly horizontal correlations between the detected chemical abundances and cloud properties. This is enabled by the clear sector of the terminator, which allows POSEIDON to disentangle the reference pressure from that of the cloud deck and hence break the degeneracy between uniform clouds and chemistry. To reiterate: \emph{non-uniform terminator cloud coverage enables precise determination of chemical abundances from transmission spectra}.

Having verified POSEIDON's ability to correctly retrieve atmospheric properties from high-precision transmission spectra, we now turn our attention towards the observed transmission spectrum of HD 209458b.

\section{Results: Application to HD 209458b}\label{section:results}

We here report the first application of POSEIDON to an exoplanet transmission spectrum. Specifically, we consider the visible \& near-infrared spectrum of HD 209458b presented in \citet{Sing2016}, as this is the highest quality transmission spectrum presently available. Our retrieval, which includes observations over the range $\sim$ $0.3-1.7 \micron$ taken by the HST STIS and WFC3 instruments, is the most extensive to date, involving the computation of an unprecedented $10^{8}$ spectra. 

We ran a comprehensive suite of retrievals spanning the model space of cloudy atmospheres. Four different cloud/haze prescriptions were considered: partial cloud coverage (both with and without hazes), uniform clouds and cloud-free atmospheres. Within each cloud prescription we performed nested model comparisons, whereby multiple retrievals are performed with chemical species selectively removed to evaluate their detection significances. This amounts to 8 independent retrievals, with $\sim$ 5 $\times$ $10^6$ model computations each, for each cloud prescription, i.e. $\gtrsim$$10^8$ model computations in total. We finally report constraints on the terminator chemical abundances and temperature profile for the cloud model most preferred by the data. We additionally illuminate how the assumed cloud model influences the inferred chemical abundances.

\subsection{Cloud Properties}\label{subsection:results_cloud_properties}

\begin{table}
\ra{1.3}
\caption{Bayesian Model Comparison of the Terminator Cloud $\hspace{7em}$ Distribution on HD 209458b}
\begin{tabular*}{\columnwidth}{l@{\extracolsep{\fill}} cccccl@{}}\toprule
$\mathrm{Model}$ & \multicolumn{1}{p{1cm}}{\centering \hspace{-0.4cm} Evidence \\ \centering $ \hspace{-0.2cm} \mathrm{ln}\left(\mathcal{Z}_{i}\right)$}  & \multicolumn{1}{p{1cm}}{\centering Best-fit \\ \centering $ \chi_{r, \mathrm{min}}^{2}$} & \multicolumn{1}{p{1.7cm}}{\centering \hspace{-0.3cm} Bayes Factor \\ \centering $ \hspace{-0.1cm} \mathcal{B}_{0i}$}& \multicolumn{1}{p{1cm}}{\centering \hspace{-0.4cm} Detection \\ \centering \hspace{-0.3cm} of Ref.}\\ \midrule
Patchy Clouds & $ 953.16 $ & $ 1.45$ & Ref. & Ref.\\
Uniform Clouds & $ 944.91 $ & $ 1.52$ & $3.8 \times 10^{\, 3}$ & $\boldsymbol{4.5 \sigma}$\\
Cloud Free & $ 940.47 $ & $ 1.62$ & $3.3 \times 10^{\, 5}$ & $\boldsymbol{5.4 \sigma}$\\
No Haze & $ 949.57 $ & $ 1.57$ & $36$ & $\boldsymbol{3.2 \sigma}$\\
Fixed Fraction & $ 953.60 $ & $ 1.44$ & $0.65$ & N/A\\
\bottomrule
\vspace{0.1pt}
\end{tabular*}
$\textbf{Notes}:$ The `fixed fraction' model has $\bar{\phi}$ held at the best fit value from the `patchy clouds' model (0.47). The `no haze' model is a patchy cloud model that considers only $\mathrm{H}_2$-Rayleigh scattering in cloudy regions. $\chi_{r, \mathrm{min}}^{2}$ is the minimum reduced chi-square.  An $n \sigma$ detection ($n \geq 3$) indicates the degree of preference for the reference model over the alternative model.
\label{table:cloud_models}
\end{table}

\begin{figure}
	\includegraphics[width=\columnwidth, trim={-0.1cm 0.1cm -0.1cm 0.3cm},clip]{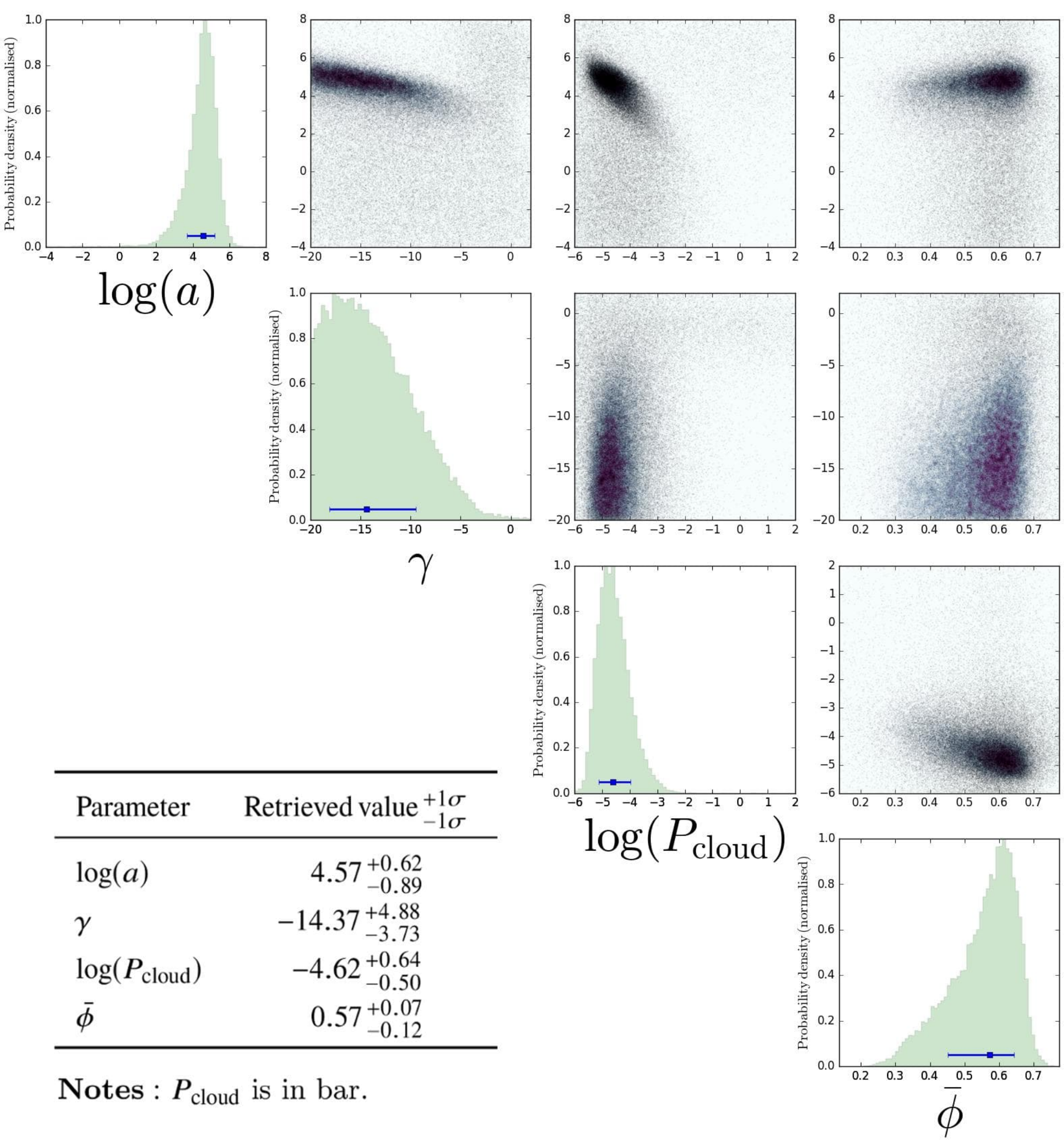}
    \caption{HD 209458b's terminator cloud properties, retrieved within the `patchy clouds' model (see Table \ref{table:cloud_models}). $\textbf{Main figure:}$ corner plot depicting correlations between pairs of derived parameters and marginalised histograms for the values of each parameter extracted by the retrieval. $\textbf{Table inset:}$ median retrieved cloud parameters and $1\sigma$ confidence levels.}
    \label{fig:posterior_clouds}
\end{figure}

\begin{figure*}
	\includegraphics[width=\textwidth, trim={0.5cm .6cm 0.5cm 0.2cm},clip]{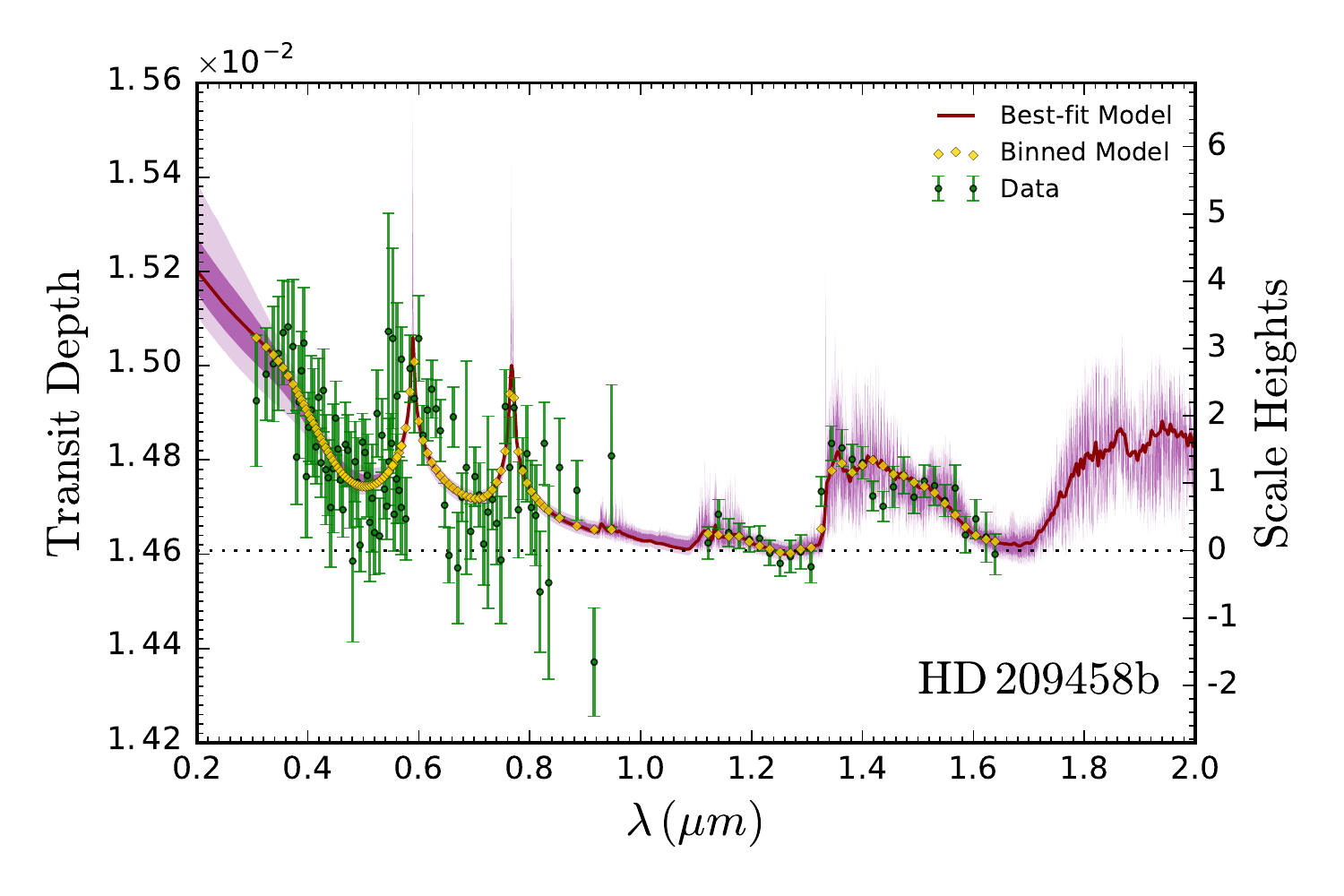}
    \caption{POSEIDON's retrieval of the visible \& near-infrared transmission spectrum of HD 209458b. The observed transit depth is indicated by green circles with error bars. The dark red curve is a Gaussian smoothed representations of the best-fit retrieved spectrum. The yellow diamonds are the median binned model points produced by the retrieval. The dark and light purple regions indicate $1\sigma$ and $2\sigma$ confidence regions (at $R \approx 10000$) in the transit depth at each wavelength derived from 10,000 random sample draws from the posterior distribution. The black dotted line indicates the value of $(R_{p}/R_{*})^{\, 2}$. The number of equivalent scale heights above this reference baseline is computed with respect to the median retrieved photosphere temperature (see section \ref{subsection:results_temperature_structure}).}
    \label{fig:Sing_spectrum}
\end{figure*}

\begin{figure*}
	\includegraphics[width=\linewidth, trim={-0.35cm .05cm -0.05cm -0.5cm},clip]{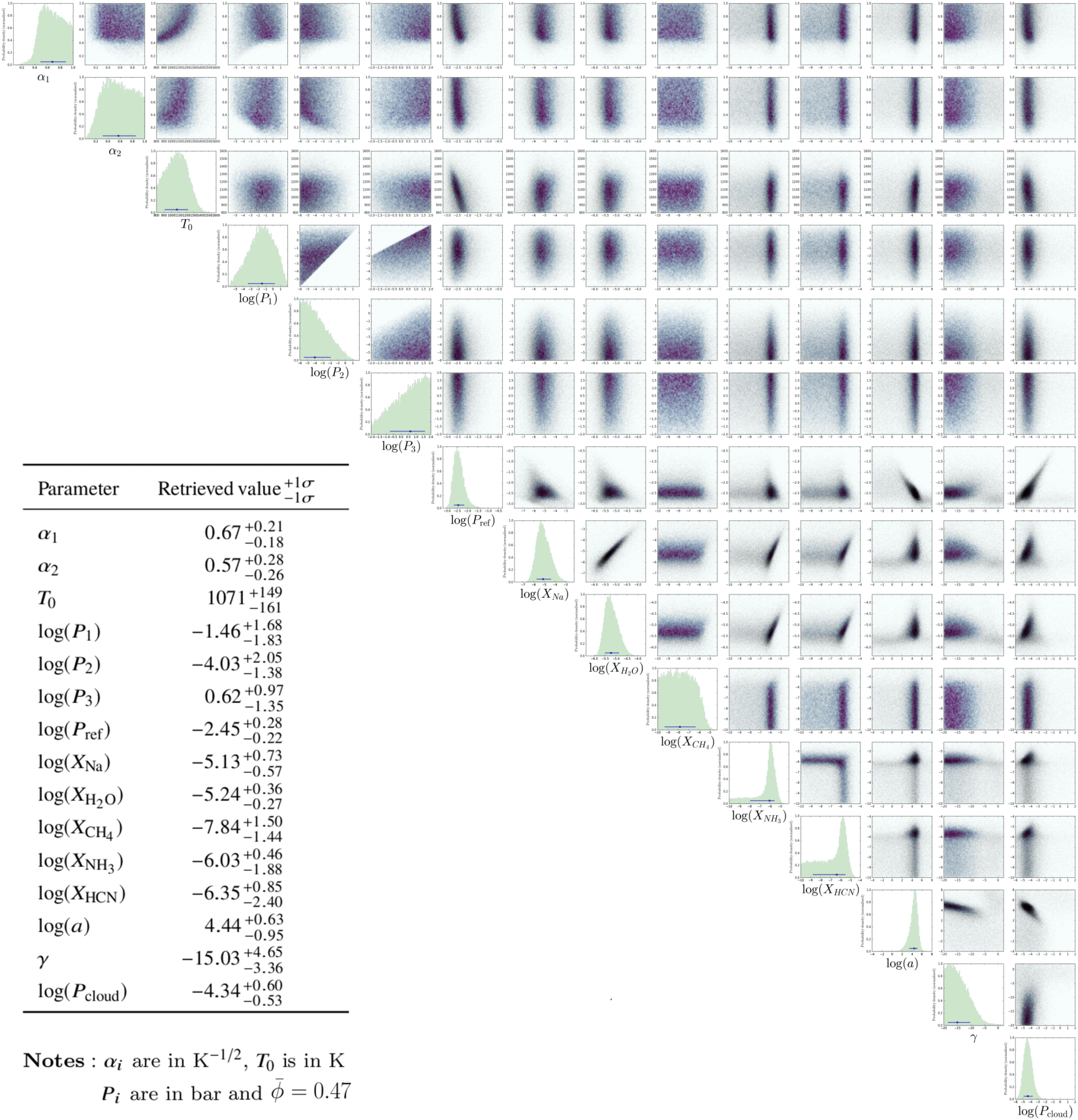}
    \caption{Full posterior distribution from POSEIDON's retrieval of the transmission spectrum of HD 209458b. $\textbf{Main figure:}$ corner plot depicting correlations between pairs of retrieved parameters and marginalised histograms for the values of each parameter extracted by the retrieval. The abundances of $\mathrm{Na}$ and $\mathrm{H_{2}O}$ are tightly constrained, relatively independently of the clouds parameters -- vindicating the prediction from the synthetic data retrieval shown in Figure \ref{fig:sim_posterior} and discussed in section \ref{subsubsection:cloud_degeneracy}. $\textbf{Table inset:}$ median retrieved values and $1\sigma$ confidence levels for each parameter, following marginalisation over the other 14 dimensions of the parameter space.}
    \label{fig:Sing_posterior}
\end{figure*}

\begin{figure*}
	\includegraphics[width=\textwidth, trim={0.2cm 0.5cm 0.0cm 0.0cm},clip]{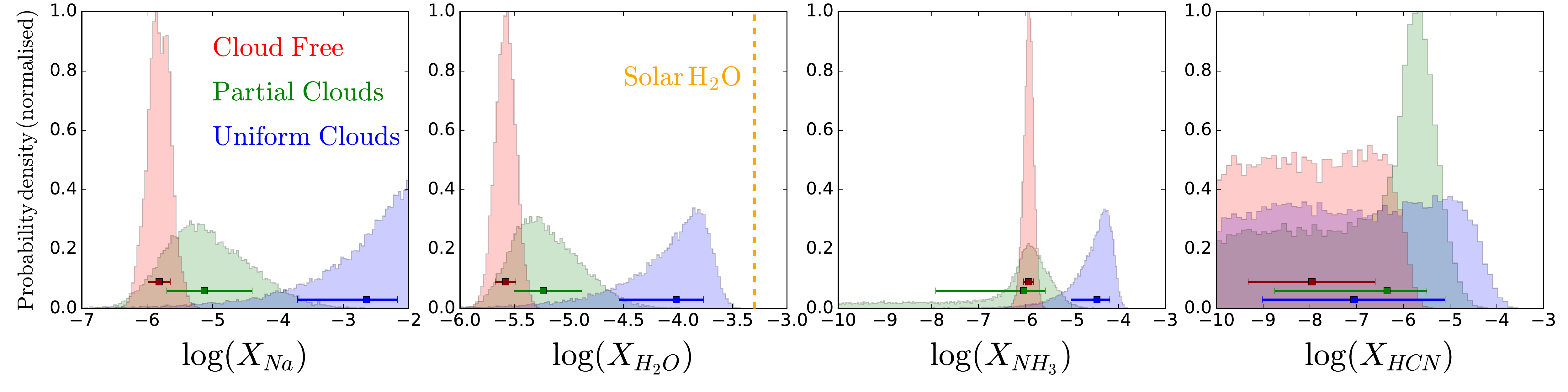}
    \caption{Marginalised posterior distributions of the chemical abundances on the terminator of HD 209458b. The red, green and blue histograms are inferences from the `cloud-free', `fixed fraction' and `uniform clouds' models summarised in Table \ref{table:cloud_models}. $\mathrm{H_{2}O}$, $\mathrm{Na \, / \, K}$ and $\mathrm{NH_3 \, / \, HCN}$ are detected in all three models (Table \ref{table:chemistry_models}). $\mathrm{NH_3}$ is detected in both the cloud-free and uniform cloud models at confidence levels of $4.9\sigma$ and $3.3\sigma$ respectively. All three models are inconsistent with a solar water abundance (indicated by the dashed orange line at $\mathrm{log}(X_{\mathrm{H_{2}O}})$ = -3.3) at $>3\sigma$ confidence. The uniform cloud model is biased to higher abundances (see text for discussion), whilst the cloud-free model is consistent with the abundances from the preferred partial cloud model. The cloud-free model underestimates the uncertainty in the derived abundances. For clarity, we do not show the posterior of the `patchy clouds' model of Table \ref{table:cloud_models} (which is almost identical to that of the `fixed fraction' model). We also do not show $\mathrm{CH_4}$, as it is unconstrained in all our retrievals (e.g. Figure. \ref{fig:Sing_posterior}).}
    \label{fig:results_chemistry}
\end{figure*}

We detect the presence of partial cloud coverage across the terminator of HD 209458b. Table \ref{table:cloud_models} summarises the results of our Bayesian model comparison, which indicates a $4.5\sigma$ preference for the patchy cloud model over the uniform cloud model. The cloud-free model is ruled out to $>5\sigma$ when compared to the patchy cloud model and to $3.4\sigma$ when compared to uniform clouds. These detection significances can be understood by examining the posterior of the patchy cloud retrieval (Figure \ref{fig:posterior_clouds}), which indicates a cloud fraction of $\bar{\phi} = 0.57^{\, +0.07}_{\, -0.12}$. Given that the cloud fraction is closer to 1 than 0, it is unsurprising that the Bayesian evidence of the uniform cloud model exceeds that of the cloud-free model. Even after marginalising over the other parameters, the posterior probability distribution of $\bar{\phi}$ is inconsistent with values of 0 or 1 ($5\sigma$ range: $0.14-0.77$); reinforcing the large penalty the Bayesian evidence suffers when forced to consider models fixed at these values. As one final assessment of the presence of partial clouds, we conducted an additional retrieval where the cloud fraction was held fixed at the best fit (min $\chi_{r}^{2}$) value from the patchy cloud retrieval ($\bar{\phi} \approx 0.47$). This model, the `fixed fraction' model in Table \ref{table:cloud_models}, possess the largest value of the Bayesian evidence amongst our cloud models and hence reinforces our assertion that patchy clouds are favoured by the data.

The detection significance of patchy clouds is found to be somewhat sensitive to the lower limit of the temperature prior considered. Identical retrievals with a lower limit on $T_0$ of 400 K result in values of $\mathrm{ln} \, \mathcal{Z}$ of 949.35 and 947.17 for uniform clouds and cloud-free atmospheres respectively. The corresponding detection significances for  patchy clouds are 3.2 $\sigma$ and 3.8 $\sigma$, respectively. This effect is caused by a tendency for both cloud-free and uniform cloud models to favour lower temperature solutions \citep[e.g., see][]{Tsiaras2016a}

We infer a high altitude ($\sim 0.01-0.1$ mbar) cloud deck on the cloudy fraction of the terminator. Above this cloud deck, we report a moderate detection ($3.2\sigma$) of high-altitude hazes. The necessity of strongly enhanced Rayleigh scattering ($\sim 5,000-100,000 \times \, \mathrm{H_{2}}$-Rayleigh) is visibly apparent in Figure \ref{fig:Sing_spectrum} from the steep increase in the transit depth towards shorter wavelengths. In addition to the high strength of the scattering coefficient, the generalised slope is remarkably negative, tending to prefer values towards the lower edge of the prior. The presence of such a strong scattering cross section at these altitudes suggests two immediate possibilities: i) incredibly light particles capable of being lofted to these altitudes by vertical mixing; ii) continuous replenishment of the species at altitude (e.g. by photochemical reactions).

In what follows, we select the `fixed fraction' cloud model, unless otherwise stated. This is the model that maximises the Bayesian evidence, and thus holds the greatest sway in light of the data. We display the full posterior of this model in Figure \ref{fig:Sing_posterior}. Note in particular that the values of the cloud parameters (and their associated errors) in this model remain consistent to within $1\sigma$ of the values shown in Figure \ref{fig:posterior_clouds}. This indicates that the uncertainty induced by allowing the cloud fraction to vary as a free parameter does not overly effect the inferences of the remaining cloud parameters compared to the case when it is fixed at the best fit value. In order to illustrate the importance of selecting the cloud model that is supported by the data, in the next section we will demonstrate how the inferred values and constraints on the retrieved chemical abundances crucially depend on the assumed cloud model.

\subsection{Chemistry}\label{subsection:results_chemistry}

\subsubsection{Detections}\label{subsubsection:results_detections}

We confirm previous detections of $\mathrm{Na}$ \citep{Charbonneau2002,Snellen2008} and $\mathrm{H_{2}O}$ \citep{Deming2013} in the transmission spectrum of HD 209458b. Our nested model comparison establishes the presence of $\mathrm{H_{2}O}$ and an alkali absorber ($\mathrm{Na \, / \, K}$) at $9.1\sigma$ and $7.3\sigma$ confidence, respectively (Table \ref{table:chemistry_models}). We do not detect $\mathrm{CH_4}$, though we establish at $>10\sigma$ confidence that the presence of either $\mathrm{CH_4}$ or $\mathrm{H_{2}O}$ is required \citep[due to their overlapping absorption features at 1.15 \micron \, and 1.40 \micron \, -- see][and Figure \ref{fig:cross_sections}]{Benneke2013}.

We additionally detect the presence of nitrogen chemistry (in the form of $\mathrm{NH_3}$ and/or $\mathrm{HCN}$) at $3.7\sigma$ confidence. The Bayes factor of the model including nitrogen chemistry compared to the model without is 186 - indicating `strong evidence' in favour of the presence of $\mathrm{NH_3}$ and/or $\mathrm{HCN}$ on the Jeffreys' scale. This detection is robust to the assumed cloud model, rising to $4.9\sigma$ and $7.7\sigma$ in uniformly cloudy and cloud-free models, respectively. In all cases, the chemical detections and confidences are insensitive to the lower limit on the temperature prior.

\begin{figure*}
	\includegraphics[width=\textwidth, trim={0.5cm 0.6cm 0.5cm 0.25cm},clip]{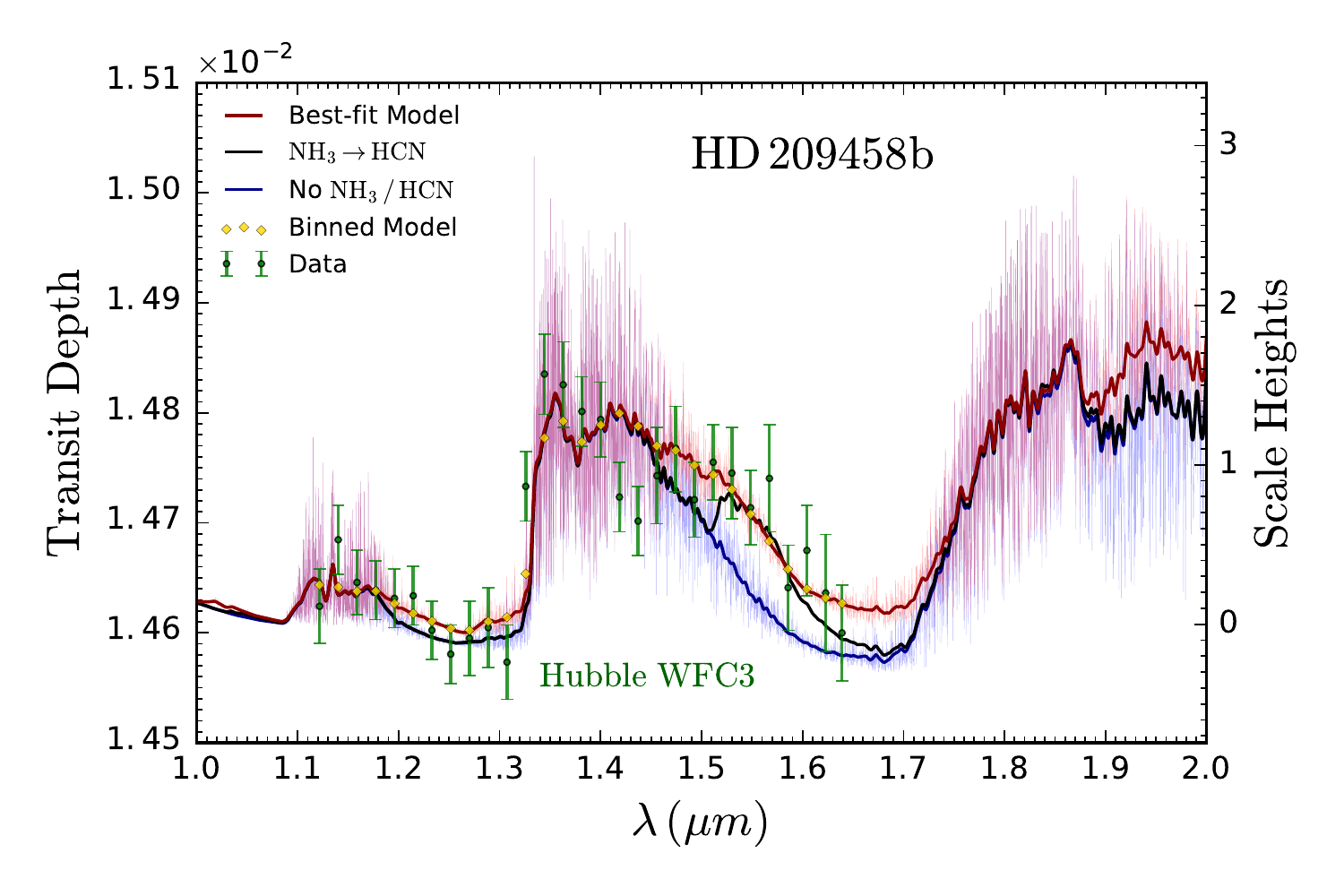}
    \caption{Evidence of nitrogen chemistry on the terminator of HD 209458b. The observed transit depth is indicated by green circles with error bars. The best fit spectrum (2ppm $\mathrm{NH_3}$) is shown in red at a resolution of $R \approx 10000$. The yellow diamonds are the binned model points corresponding to the best-fit spectrum. The black spectrum is a model identical to the best fit with the abundances of $\mathrm{NH_3}$ and $\mathrm{HCN}$ interchanged (such that $\mathrm{HCN}$ becomes the dominant nitrogen-bearing species). The blue spectrum is a model identical to the best fit with $\mathrm{NH_3}$ and $\mathrm{HCN}$ removed. The dark red, black and blue curves are Gaussian smoothed representations of the high-resolution spectra with corresponding colours. Performing full atmospheric retrievals, exploring our entire parameter space of chemistry, temperature structure and clouds/hazes, establishes that the model including $\mathrm{NH_3}$ and $\mathrm{HCN}$ (red) is preferred by the data over models with no nitrogen chemistry at $3.7\sigma$ confidence. The primary evidence for a nitrogen-bearing species comes from the additional absorption over the spectral range $\sim 1.45-1.70 \,\micron$.}
    \label{fig:nitrogen_chemistry}
\end{figure*}

When considering the partial cloud model preferred by the data (Table \ref{table:cloud_models}), we are unable to distinguish between the presence of $\mathrm{NH_3}$ and $\mathrm{HCN}$. This is due to the effect of partial clouds in altering the slopes of absorption features (see Figure \ref{fig:cloud_parameters}). However, in both cloud-free and uniformly cloudy models this degeneracy  is lifted and $\mathrm{NH_3}$ is detected at $4.9\sigma$ ($\mathcal{B}_{0i}$ = 22,000) and $3.3\sigma$ ($\mathcal{B}_{0i}$ = 58) confidence respectively. We do not detect $\mathrm{HCN}$ in the cloud-free or uniformly cloudy models, obtaining flat abundance posteriors with established upper limits of $\sim 10^{-6}$ and $10^{-4}$ respectively. The influence of nitrogen chemistry on the transmission spectrum is shown in Figure \ref{fig:nitrogen_chemistry} and discussed further in section \ref{subsubsection:nitrogen_chemistry}.

\begin{table}
\ra{1.3}
\caption[]{Bayesian Model Comparison of the Chemistry on the Terminator of HD 209458b}
\begin{tabular*}{\columnwidth}{l@{\extracolsep{\fill}} cccccl@{}}\toprule
$\mathrm{Model}$ & \multicolumn{1}{p{1cm}}{\centering \hspace{-0.4cm} Evidence \\ \centering $ \hspace{-0.2cm} \mathrm{ln}\left(\mathcal{Z}_{i}\right)$}  & \multicolumn{1}{p{1cm}}{\centering Best-fit \\ \centering $ \chi_{r, \mathrm{min}}^{2}$} & \multicolumn{1}{p{1.7cm}}{\centering \hspace{-0.3cm} Bayes Factor \\ \centering $ \hspace{-0.2cm} \mathcal{B}_{0i}$}& \multicolumn{1}{p{1cm}}{\centering \hspace{-0.4cm} Detection \\ \centering \hspace{-0.2cm} of Ref.}\\ \midrule
Full Chemistry & $ 953.60 $ & $ 1.44 $ & Ref. & Ref.\\
No $\mathrm{H_2{O} \, / \, CH_4}$ & $ 904.62 $ & $ 2.35 $ & $ 1.9 \times 10^{\, 21} $ & \hspace{-0.5em} $\boldsymbol{10.1 \sigma}$ \\
No $\mathrm{H_{2}O}$ & $ 914.62 $ & $ 2.14 $ & $ 8.4 \times 10^{\, 16} $ & $\boldsymbol{9.1 \sigma}$ \\
No $\mathrm{Na \, / \, K}$ & $ 928.92 $ & $ 1.93 $ & $ 5.2 \times 10^{\, 10} $ & $\boldsymbol{7.3 \sigma}$ \\
No $\mathrm{NH_3 \, / \, HCN}$ & $ 948.37 $ & $ 1.53 $ & $ 186 $ & $\boldsymbol{3.7 \sigma}$ \\
No $\mathrm{NH_3}$ & $ 952.80 $ & $ 1.44 $ & $2.2 $ & N/A \\
No $\mathrm{HCN}$ & $ 953.35 $ & $ 1.42 $ & $1.3 $ & N/A \\
No $\mathrm{CH_4}$ & $ 954.01 $ & $ 1.42 $ & $0.7 $ & N/A \\
\bottomrule
\vspace{0.1pt}
\end{tabular*}
$\textbf{Notes}:$ The `full chemistry' model includes opacity due to $\mathrm{H_2}, \,  \mathrm{He}, \, \,  \mathrm{Na}, \,  \mathrm{K}, \,  \mathrm{H_{2}O}, \,  \mathrm{CH_{4}}, \,  \mathrm{NH_{3}}, \, \mathrm{HCN}$ and corresponds to the `fixed fraction' cloud model given in Table \ref{table:cloud_models}. $\chi_{r, \mathrm{min}}^{2}$ is the minimum reduced chi-square.  An $n \sigma$ detection ($n \geq 3$) indicates the degree of preference for the reference model over the alternative model.
\label{table:chemistry_models}
\end{table}

\subsubsection{Abundance Constraints}\label{subsubsection:results_abundance}

We report our constraints on the chemical abundances on HD 209458b's terminator in Figure \ref{fig:results_chemistry}. The abundances we report are amongst the most precise ever obtained from an exoplanet transmission spectrum ($\sim 0.3$ dex for $\mathrm{H_{2}O}$). This is despite our marginalisation over two-dimensional terminator cloud coverage, due to POSEIDON's ability to break the degeneracy between clouds and chemistry (Section \ref{subsubsection:cloud_degeneracy}). We explore how the assumed cloud distribution affects the inferred abundances in section \ref{subsubsection:influence_of_clouds}.

The terminator of HD 209458b is inconsistent with a solar $\mathrm{H_{2}O}$ abundance. This is established at $>5\sigma$ confidence for both partial cloud and cloud-free models and at $>3\sigma$ confidence for uniform clouds. The retrieved value, $\mathrm{log}(X_{\mathrm{H_{2}O}}) = -5.24^{\, +0.36}_{\, -0.27}$, is remarkably consistent with the values reported by both \citet{Madhusudhan2014c} $\left(-5.27^{\, +0.65}_{\, -0.16}\right)$ and \citet{Barstow2016} ($-5.3$ to $-5.0$). This is unsurprising, as the observed spectrum (Figure \ref{fig:Sing_spectrum}) shows the amplitude of the H$_2$O absorption feature at $1.4 \, \micron$ is only 2 scale-heights -- whereas a solar composition atmosphere at a similar temperature and cloud coverage fraction would give $\sim$ 5 scale heights (Figure \ref{fig:sim_retrieved_spectrum}).

We demonstrate that the $\mathrm{Na}$ abundance can be reasonably well-constrained ($\sim 0.6$ dex), despite the $\sim 120$ ppm errors in the optical STIS data. Whilst this serves as an important demonstration of principle, we caution against reading too much into the retrieved values. This is due to the simplicity of our treatment of the alkali cross sections in the present work (see section \ref{subsubsec:chemistry}). We will address precise alkali abundance constraints in future work.

The abundances of both $\mathrm{NH_3}$ and $\mathrm{HCN}$ show a sharp peak at $\sim 10^{-6}$ with a tail towards lower abundances. The abundance of ammonia is the best constrained at $\mathrm{log}(X_{\mathrm{NH_{3}}}) = -6.03^{\, +0.46}_{\, -1.88}$ ($0.01-2.7$ ppm). The tails stem from the fact that either of these nitrogen-bearing species can explain the observed absorption features - if one has high abundance, the other will have low abundance and vice versa. Ultimately it is this long tail that prevents a unique determination of the species causing the absorption. It can be seen from the lower probability density of the tail in Figure \ref{fig:results_chemistry} (and the higher Bayes factor in Table \ref{table:chemistry_models}) that the presence of $\mathrm{NH_3}$ is marginally preferred over $\mathrm{HCN}$ when considering partial cloud coverage. This symmetry is broken when considering cloud-free or uniformly cloudy models, both of which feature well-constrained $\mathrm{NH_3}$ and a flat posterior for $\mathrm{HCN}$ - explaining the $\mathrm{NH_3}$ detections observed in these models. Given the combination of these high $\mathrm{NH_3}$ detection significances and the coincident peak of its abundance distribution between the cloud-free and the partial cloud models, we strongly suspect that it is this species, not $\mathrm{HCN}$, that is the source of the detected nitrogen chemistry.

\subsubsection{Nitrogen Chemistry}\label{subsubsection:nitrogen_chemistry}

We now proceed to identify the absorption features giving rise to our detection of nitrogen chemistry. In Figure \ref{fig:nitrogen_chemistry} we show the effect on our best-fit spectrum (red) $\left(\mathrm{log} \, X_{\mathrm{\left[H_{2}O, \, CH_{4}, \, NH_{3}, \, HCN\right]}} = [-5.21, \, -8.63, \, -5.72, \, -8.39]\right)$ of removing the $\mathrm{NH_3}$ and $\mathrm{HCN}$. Given that $\mathrm{NH_3}$ is the dominant nitrogen-bearing molecule for this spectrum, this amounts to an assessment of the impact of ammonia on near-infrared transmission spectra.

The primary impact of $\mathrm{NH_3}$ absorption in the WFC3 bandpass is to raise the transit depth of HD 209458b by $\sim 5 \times 10^{-5}$ compared to what would be expected from pure $\mathrm{H_{2}O}$ absorption (blue) over the spectral range $\sim 1.45-1.7 \,\micron$. A secondary feature of magnitude $\sim 1 \times 10^{-5}$ is seen between  $\sim 1.2-1.3 \,\micron$. These absorption features are readily identified by an examination of the $\mathrm{NH_3}$ cross section (Figure \ref{fig:cross_sections}), which is seen to dominate that of $\mathrm{H_{2}O}$ over these two regions. The necessity of additional absorption is evident from the data itself, as there are 4 data points elevated by $2\sigma$ and one point elevated by $3\sigma$ above the model without nitrogen chemistry.

We now offer suggestions on how to distinguish between $\mathrm{NH_3}$ and $\mathrm{HCN}$ in transmission spectra when nitrogen chemistry is detected. The difficulty inherent in this task is demonstrated in Figure \ref{fig:nitrogen_chemistry}, where the black curve shows the effect of interchanging the abundances of $\mathrm{NH_3}$ and $\mathrm{HCN}$, such that $\mathrm{HCN}$ becomes the dominant nitrogen-bearing molecule. $\mathrm{HCN}$ causes an increase in the transit depth that is almost identical to that of $\mathrm{NH_3}$ from $\sim 1.53-1.6 \,\micron$ (Figure \ref{fig:cross_sections}) , though it generally matches pure $\mathrm{H_{2}O}$ absorption outside this range. The degeneracy with $\mathrm{NH_3}$ may be lifted by high resolution observations in three regions of the WFC3 bandpass: i) $1.2-1.32 \,\micron$; ii) $1.46-1.52 \,\micron$; iii) $1.6-1.7 \,\micron$. Given that the difference between $\mathrm{NH_3}$ and $\mathrm{HCN}$ dominated spectra are of the order of the error bars (35ppm), this is pushing the frontier of current observational capabilities.

There is, however, another potential avenue that may enable the unique detection of $\mathrm{NH_3}$ / $\mathrm{HCN}$ with WFC3 observations. Namely, the sharp $\mathrm{NH_3}$ posterior shown in Figure \ref{fig:results_chemistry}'s cloud-free model suggests that genuinely cloud-free atmospheres may allow highly robust detections of $\mathrm{NH_3}$ and/or $\mathrm{HCN}$. We discuss this possibility further in what follows.

\subsubsection{The Influence of Clouds on Chemical Abundances}\label{subsubsection:influence_of_clouds}

We now proceed to quantify the extent to which the cloud model assumed by a retrieval can influence the inferred chemical abundances. We have already identified an innate challenge in distinguishing between $\mathrm{NH_3}$ and $\mathrm{HCN}$ in partially cloudy transmission spectra. We further saw in section \ref{subsubsection:results_detections} and Figure \ref{fig:results_chemistry} that the detection significances and abundance constraints depend crucially on the cloud model employed by retrievals.

In general, our cloud-free abundances are artificially well constrained at lower values than those inferred by the preferred partial cloud model (though they remain consistent within $1\sigma$). However, this suggests that planets with genuinely low cloud coverage may permit strong detections with precise abundance constraints. This is especially evident in the case of nitrogen chemistry, where our cloud-free model clearly identifies the presence of $\mathrm{NH_3}$ at $\approx5\sigma$ and constrains its abundance to $\mathrm{log}(X_{\mathrm{NH_{3}}}) = -5.92^{\, +0.10}_{\, -0.11}$. We hence suggest that, even given current observational errors, precise abundance determinations of nitrogen-bearing molecules may be obtained on planets with low overall terminator cloud coverage.

In contrast, the uniform cloud model tends to overestimate chemical abundances. This is strikingly apparent in the case of $\mathrm{Na}$, which favours unphysical values towards the upper limit of the prior ($10^{-2}$). The biasing of abundances to erroneously high values under the assumption of uniform clouds is a consequence of a fundamental degeneracy between clouds, hazes, the reference pressure and the chemical abundances. For uniform terminator cloud coverage, the cloud deck pressure, $P_{\mathrm{cloud}}$ and the pressure at the radius of the planet, $P_{\mathrm{ref}}$, are equal, existing on a line in parameter space without a unique solution. Defining $P_{\mathrm{ref}}$ at a radii different to $R_p$ merely offsets the line of degeneracy. This family of solutions determines the baseline of the spectrum (Figure \ref{fig:Sing_spectrum}, black dotted line). When $P_{\mathrm{ref}}$ and $P_{\mathrm{cloud}}$ are lowered along this solution line, both the amplitude of spectral features and the Rayleigh-enhancement factor can rise to produce an identical spectrum. We have verified that this biasing to higher abundances is an artefact of the uniform cloud model itself by running simulated retrievals with solar $\mathrm{Na}$ abundances and confirming that the behaviour seen in Figure \ref{fig:results_chemistry} can be reproduced for synthetic data. We thus caution against the blind application of uniform cloud models; indeed, our results suggest that cloud-free models are a better option if one solely wishes to estimate the chemical abundances of a transiting exoplanet to within $1\sigma$

We have shown that partial cloud coverage breaks the degeneracy between clouds and chemistry imposed artificially by the assumption of uniform clouds (section \ref{subsection:retrieval_validation}). Alternatively, this degeneracy may be broken by assuming a priori knowledge of the scattering slope in the optical \citep[e.g][]{Benneke2012} or of the molecular/condensate chemistry \citep{Benneke2015}. The attraction of partial clouds is that we do not have to make such assumptions, allowing us to directly retrieve cloud, haze, chemical and temperature properties of the atmosphere simultaneously. Though we have shown that partial clouds enable precise determination of the $\mathrm{H_{2}O}$ and  $\mathrm{CH_{4}}$ abundances, and are preferred by the Bayesian evidence at high significance, they complicate the interpretation of nitrogen chemistry as they tend to render the slopes of molecular features more shallow (Figure \ref{fig:cloud_parameters}, bottom right). This decreases the magnitude of the transit depth `gap' induced by nitrogen chemistry over the range $\sim 1.45-1.70 \,\micron$ (Figure \ref{fig:nitrogen_chemistry}) by broadening the width of the pure $\mathrm{H_{2}O}$ absorption feature.

Given these chemical and cloud inferences, we now proceed to present our retrieved P-T profile of the terminator of HD 209458b. The combination of all three of these properties in required to build a coherent picture of the conditions on the terminator.

\subsection{Temperature Structure}\label{subsection:results_temperature_structure}

\begin{figure}
	\includegraphics[width=\columnwidth, trim={-0.07cm 0.7cm 0.0cm 0.35cm},clip]{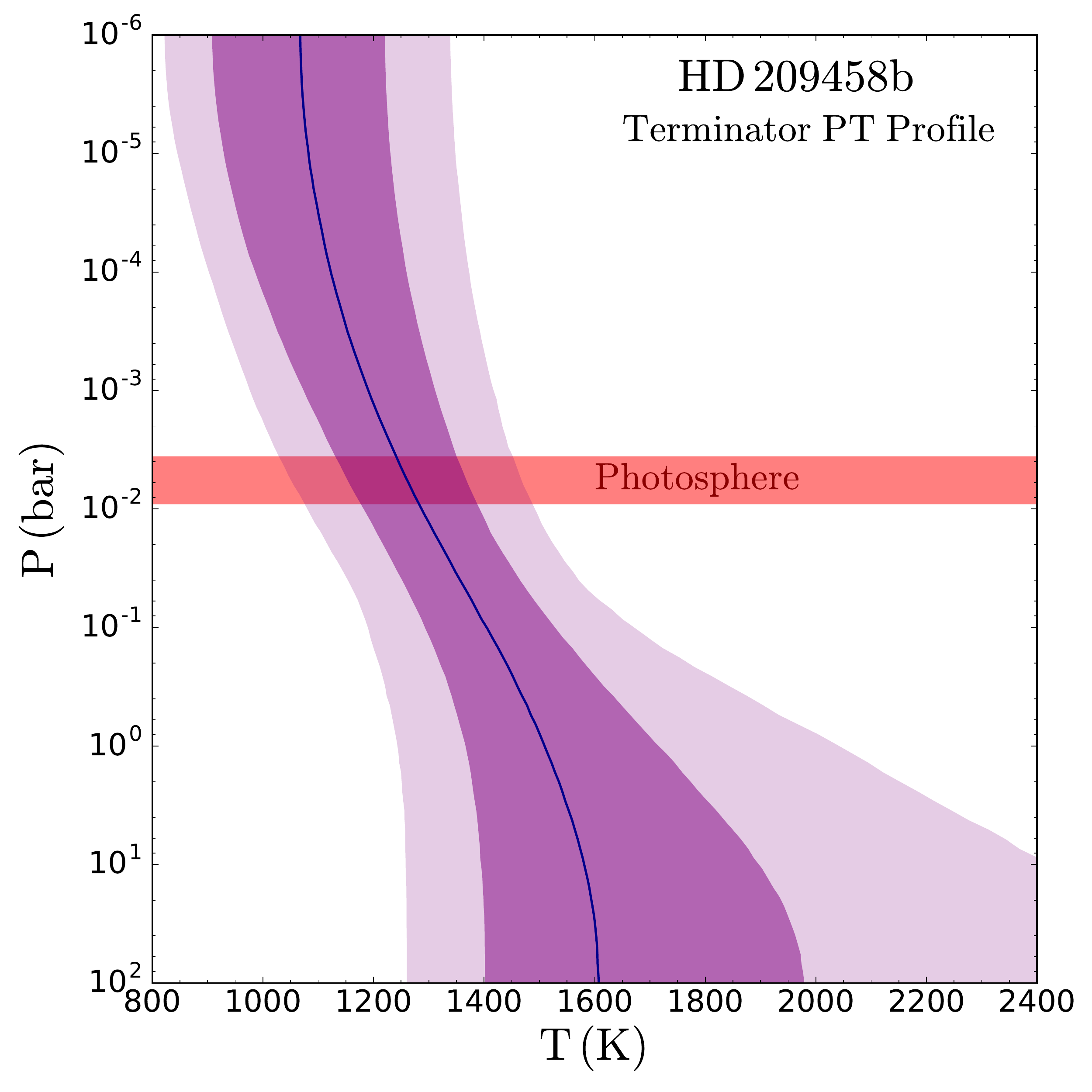}
    \caption{POSEIDON's retrieval of the terminator P-T profile on HD 209458b. The dark blue curve represents the median P-T profile, with the dark and light purple regions indicating the $1\sigma$ and $2\sigma$ confidence regions of the temperature in each layer (derived from 10,000 random sample draws from the posterior). The  $1\sigma$ extent of the near-infrared photosphere ($\tau = 1$ at $1.5 \micron$) is shown by the red shaded region. The median temperature in the photosphere on the terminator, 1221 K, is approximately 200 K below the planetary equilibrium temperature, with the temperature changing by $\sim 50$ K across the photosphere.}
    \label{fig:Sing_PT}
\end{figure}

The temperature structure on the terminator of HD 209458b is not isothermal. If the data supported an isothermal profile, we would expect to see $\alpha_{1, \,2}$ tending towards larger values, which we do not observe (see Figure \ref{fig:Sing_posterior}). Whilst unsurprising from physical arguments and Global Circulation Model (GCM) simulations, this point merits emphasis. It is often assumed in transmission retrieval that with currently available data: i) it is not possible to retrieve the shape of the terminator P-T profile; ii) an isothermal profile does not overly affect the inferred abundances. Here we demonstrate the invalidity of the first assumption for high-precision data, which was also examined by \citet{Barstow2013}, and for a critical examination of the second we refer the reader to \citet{Rocchetto2016}.

Figure \ref{fig:Sing_PT} shows our retrieved P-T profile. We highlight in red the near-infrared photosphere ($\tau = 1$ at $1.5 \micron$), as this is the region predominately probed in transmission. Notably, the temperature is not constant across the photosphere, changing by $\sim 50$ K. By assuming an isothermal profile, this behaviour, and its effect on molecular cross sections, will not be captured. As expected intuitively, a tight constraint on the temperature is obtained at these altitudes ($T_{\mathrm{phot}} = 1221^{+131}_{-138} \,$K). The confidence regions naturally expand away from the regions directly probed by the observations, particularly at pressures $\gtrsim 100$ mbar where we would usually expect the atmosphere to be opaque in slant geometry due to collisionally-induced opacity alone. Our retrieved photospheric temperature is some 200 K colder than the planetary equilibrium temperature ($T_{\mathrm{eq}} \approx 1450$ K). This is unsurprising, given that transmission spectra probe high altitudes in the cooler terminator region. Again, we emphasise that such inferences are made possible by the high-precision transmission spectrum.

This profile represents the average terminator P-T profile. Strictly speaking, we expect this to be composed of two underlying profiles: a cooler profile in the cloudy terminator region and a warmer profile in the clear region -- as condensates tend to form in cooler regions, where the P-T profile may intersect additional condensation curves than in the warmer region. This averaging also explains the relatively high-altitude photosphere, as the opaque cloud deck we infer at $P_{\mathrm{cloud}} \approx 0.01-0.1$ mbar in the cloudy region combines with the cloud-free region to determine the $\tau = 1$ surface.

\section{Summary And Discussion}\label{section:discussion}

In this work, we have established a framework for retrieving properties of transiting exoplanet atmospheres with inhomogeneous clouds. As an initial demonstration, we applied our new atmospheric retrieval algorithm POSEIDON to the visible and near-infrared transmission spectrum of the hot Jupiter HD 209458b. Our major findings are as follows:

\begin{itemize}
  \item We have found a potential avenue to break degeneracies between clouds and chemistry. Such degeneracies are artificially imposed by assuming one-dimensional cloud coverage and can be lifted by considering two-dimensional inhomogeneous cloud distributions. This enabled us to demonstrate precise determinations of the abundances of the prominent chemical species in a given spectral bandpass.
  \item We report the first detection of nitrogen chemistry in an exoplanet atmosphere -- established at $>3.7 \sigma$ confidence. Both cloud-free and uniform cloud models identify NH$_3$ as the probable cause of nitrogen-induced absorption observed over the range $1.45-1.7 \micron$. The ammonia abundance is constrained to 0.01-2.7 ppm.
  \item The H$_2$O abundance on the terminator of HD 209458b is $30-100 \times$ sub-solar (5-15 ppm). This is established by a fully Bayesian exploration of $\sim 10^8$ transmission spectra, including two-dimensional cloud/haze distributions.
  \item Partially cloudy models are favoured by a Bayesian model comparison over both uniformly cloudy ($4.5\sigma$) and cloud-free ($5.4\sigma$) models. The terminator cloud fraction is constrained to $57^{+7}_{-12}$ per cent.
  \item Scattering due to high-altitude ($P< 0.05$ mbar) hazes is detected at $3.2\sigma$.
  \item The terminator temperature structure of a transiting exoplanet can be constrained using high-precision HST transmission spectra. The temperature in the line of sight near-infrared photosphere is constrained to $1221^{+131}_{-138}$ K. 
\end{itemize}

We have demonstrated that a wealth of information may be extracted from currently available high-precision HST transmission spectra. Our most interesting result is that non-uniform terminator properties may provide an avenue to extract \emph{additional} information from exoplanet transmission spectra. Indeed, it appears that the consideration of inhomogeneous cloud coverage may have represented `missing physics' that enabled us to break degeneracies in an otherwise underspecified problem. Exoplanet atmospheres are inherently multi-dimensional, and to treat them as one-dimensional will at best miss key insights and at worse incur unnecessary degeneracies.

\subsection{Disequilibrium Nitrogen Chemistry}

Given what is known from solar system and brown dwarf studies, our detection of nitrogen chemistry in an exoplanet atmosphere should come as no surprise. In particular, ammonia is present both on Jupiter -- in the form of high altitude clouds \citep{Sato1979} -- and in brown dwarf atmospheres \citep[e.g.][]{Roellig2004,Saumon2006,Cushing2008}. Recently, \citet{Line2015} demonstrated that NH$_3$ can be detected on brown dwarfs using low-resolution near-infrared data, though they noted a lack of obvious spectral features leading to their detection. We demonstrated in Figures \ref{fig:cross_sections} and \ref{fig:nitrogen_chemistry} that the absorption features contributing to our NH$_3$ detection in the near-infrared are located over the ranges $\sim 1.45-1.7 \,\micron$ and $\sim 1.2-1.3 \,\micron$. Additionally, the fact that \citet{Madhusudhan2016b} reported sharp NH$_3$ abundance constraints in a population of three T-type brown dwarfs using WFC3 data alone further supports our detection. We also note that the nitrogen-bearing molecule HCN has been suggested, at low statistical significance, as a potential component in the atmosphere of the super-Earth 55 Cancri e \citep{Tsiaras2016b}. However, as we have noted in section~\ref{subsubsection:results_detections} our detection significance of nitrogen chemistry varies dependant on the cloud model employed -- with the lowest confidence being that for the partial cloud model at $3.7 \sigma$. This sensitivity raises the possibility that additional physical mechanisms not considered in our models could produce a similar effect to that which we attribute to nitrogen chemistry. To adequately address this will require a concerted effort with both the development of increasingly sophisticated retrieval forward models alongside additional observations with sufficiently high precision to resolve the differences between models with and without nitrogen chemistry.

Inferring nitrogen chemistry and resulting abundance constraints heralds the opening of a new window into exoplanetary composition and atmospheric dynamics. The ammonia abundance we infer ($\approx$1 ppm) using our most preferred model represents a $\gtrsim$ 100 $\times$ enhancement over the value expected of an atmosphere in thermochemical equilibrium with solar nitrogen abundance at our constrained temperature in the photosphere \citep{Moses2011}. This suggests non-equilibrium processes may prove necessary in order to transport ammonia from regions where such high abundances can naturally form. One such avenue is transport-induced quenching, whereby regions where the characteristic dynamical timescale ($\tau_{\mathrm{dyn}}$) is less than the chemical reaction timescale ($\tau_{\mathrm{chem}}$) reflect the abundance from the `quench' level where $\tau_{\mathrm{dyn}} = \tau_{\mathrm{chem}}$. For HD 209458b, $\tau_{\mathrm{dyn}} \sim 10^5 \, \mathrm{s}$ for both horizontal and vertical advection \citep{Cooper2006}, whereas in the terminator photosphere ($P \sim 10 \, \mathrm{mbar}, T \sim 1200$ K) $\tau_{\mathrm{chem}} \sim 10^{13} \, \mathrm{s}$ for NH$_3$ $\rightarrow$ N$_2$ conversion \citep{Zahnle2014}. If horizontal quenching dominates, the NH$_3$ abundance would be expected to follow the chemical equilibrium value characteristic of the dayside, where $ \tau_{\mathrm{chem}}$ is shorter \citep{Agundez2012}. If, however, vertical quenching dominates, the NH$_3$ abundance in the observable photosphere will reflect that of the chemical equilibrium abundance at the altitude where $\tau_{\mathrm{chem}} \sim 10^{5} $, which occurs around pressures $\approx 1$bar \citep{Moses2011}. Our abundance estimate of $\approx$1 ppm is remarkably consistent with that predicted at the terminator of HD 209458b by vertical quenching models \citep{Moses2011} using nominal temperature and atmospheric mixing profiles from GCMs \citep{Showman2009} and assuming a solar abundance of nitrogen. A wider range of parameters (e.g. mixing strengths, N abundance, etc.) beyond those specifically assumed in the forward models of \citet{Moses2011} could also potentially explain the same. This suggests that NH$_3$ abundance constraints across a wide variety of exoplanets could provide a powerful diagnostic of the frequency and strength of non-equilibrium transport in a general context.

\subsection{Implications for Formation Conditions}

Recent years have seen increased interest in utilising elemental ratios as formation diagnostics. In particular, the C/O ratio is often invoked in attempts to constrain planetary formation and migration pathways relative to the snowlines of major condensates \citep[e.g.][]{Oberg2011,Madhusudhan2014b}. Nitrogen chemistry offers complimentary diagnostics, as enhanced N/H ratios are anticipated for planets forming further out in protoplanetary disks \citep{Piso2016}. Indeed, \citet{Piso2016} suggested that the N/O ratio for planets forming in outer disks could be significantly enhanced relative to the stellar value and to the C/O ratio.

Our present constraints on H$_2$O and NH$_3$ suggest a scenario where the planet formed far out in the disk and migrated to its present location by dynamical scattering. The sub-solar H$_2$O mixing ratio we find, despite the consideration of clouds/hazes, is inconsistent with an atmospheric composition of solar elemental abundances. Either a significantly sub-solar overall metallicity or super-solar C/O ratio are required to explain such low abundances \citep{Madhusudhan2014a}. On the other hand, as discussed above, the observed NH$_3$ abundance is consistent with non-equilibrium chemistry along with a nearly solar N abundance \citep{Moses2011}. Therefore, a consistent possible explanation for both the low H$_2$O and high NH$_3$ abundance we observe is the presence of a super-solar C/O as well as a super-solar N/O ratio; the metallicity can be solar in all elements except O. This composition can be achieved if the planet formed beyond the CO$_2$ or CO snowlines, accretes mostly gas  \citep{Oberg2011,Madhusudhan2014b,Piso2016} and migrated to its current orbit by disk-free mechanisms \citep{Madhusudhan2014b} or formed via pebble accretion \citep{Madhusudhan2016c}. This is further supported by the fact that the host star HD~209458 is super-solar in metallicity, including O, which means that it would be infeasible to obtain such a low oxygen abundance in the planet if it migrated through the disk and accreted planetesimals \citep[e.g.][]{Brewer2016,Mordasini2016}.

\subsection{Solar vs. Sub-solar H$_2$O Estimates}

Our robust demonstration that the terminator of HD 209458b is depleted in H$_2$O relative to solar values runs contrary to the claim asserted by \citet{Sing2016}. By not performing a retrieval, explicitly imposing thermochemical equilibrium, assuming isothermal P-T profiles and considering only global clouds/hazes their models induce sufficient a priori biases to render their conclusions unreliable. Indeed, the inadequacy of this forward model approach has already been shown by \citet{Barstow2016}, who performed a retrieval on the same dataset and found a sub-solar H$_2$O abundance ($0.01-0.02\times$ solar) in excellent agreement with ours and that of \citet{Madhusudhan2014c}. Though our H$_2$O abundances agree, the retrievals of \citet{Barstow2016} are somewhat limited by the usage of an optimal estimation algorithm, which explored only a limited volume of parameter space (3,600 models vs. our $10^{8}$) on a pre-defined grid of temperature profiles and cloud properties. Furthermore, the lack of marginalisation over parameters or Bayesian evidence computation afforded by such an algorithm renders it impractical for parameter estimation or Bayesian model comparison, such as we have conducted here. 

More sophisticated retrievals have also relied on making a priori model assumptions to break degeneracies between clouds and composition from transmission spectra. \citet{Benneke2015}, who also used a nested sampling algorithm like ours, claims a solar H$_2$O abundance on the terminator of HD 209458b. However, their approach explicitly imposed a vast array of a priori physics: P-T profiles are not retrieved (radiative-convective equilibrium is assumed), C-N-O chemical pathways are enforced and clouds, assumed to be composed of $\mathrm{Mg Si O_3}$, $\mathrm{Mg Fe Si O_4}$ and $\mathrm{Si C}$, are constructed using a model inspired by that of \citet{Ackerman2001} and are uniform across the terminator. More recently, \cite{Line2016b} attempted retrieval of the day-side atmospheric properties of HD~209458b using thermal emission spectra. They found a rather broad range in H$_2$O abundance of 0.06-10 $\times$ solar, i.e. including substantially sub-solar as well as super-solar values, at 1-$\sigma$ confidence, on the day-side. However, as they point out, the inferences are hampered by an anomalously high CO$_2$ abundance which is strongly correlated with the H$_2$O abundance. This is a well recognised problem \citep{Heng2016} in thermal emission retrievals which future work needs to investigate. 

More generally, the imposition of a priori assumptions has been used in retrievals of transmission spectra of several exoplanets where chemical and/or radiative equilibrium is enforced to derive elemental O and C abundances \citep[e.g.,][]{Benneke2015,Kreidberg2015}. These approaches, more akin to forward models, undermine the ability of a retrieval to accomplish its fundamental goal: to infer the properties of an atmosphere with an absolute \emph{minimal} set of assumptions. Succinctly put: we have shown that, in addition to clouds and hazes, a sub-solar $\mathrm{H_{2}O}$ abundance at the terminator is essential to explain the low-amplitude spectral features of HD 209458b. 

\subsection{Cloud Properties}

Our inference of a partially cloudy atmosphere along HD 209458b's terminator compliments observations of inhomogeneous clouds in both the solar system and brown dwarfs. On Earth and Jupiter, a banded cloud structure arises from atmospheric convection cells transporting gas parcels vertically, where clouds form upon crossing the relevant condensation curve, with the dry air carried to a different latitude where the formation of clouds is suppressed \citep{depater2001}. A similar mechanism has been postulated to induce latitudinally inhomogeneous clouds in brown dwarfs \citep[e.g.,][]{Marley2010}, with observational evidence also recently emerging \citep[e.g.,][]{Buenzli2012}.

Inhomogeneous cloud distributions have similarly been predicted to be common across the terminator region of hot Jupiters \citep{Parmentier2016}. The physical mechanism here is the day-night temperature contrast on tidally locked planets driving a super-rotating equatorial jet, in turn raising the temperature of the eastern terminator by hundreds of K above that of the cooler western terminator \citep{Showman2002}, where clouds are then more likely to form. Interestingly, we infer the properties of our cloudy region to consist of extremely high-altitude ($\approx 0.01-0.1$ mbar) clouds with enhanced Rayleigh scattering above the deck. The temperatures at these altitudes are less than those at which photochemical hazes are expected to form \citep[$\sim 1000-1100$ K, see][]{Zahnle2009,Moses2014}, so it is possible that the cloud deck we infer may be photochemical in origin. This possibility will require exploration by detailed photochemical models to explore its plausibility.

The clear terminator asymmetry in cloud properties on HD 209458b naturally raises the possibility of asymmetry in other observable properties. For example, \citet{Kataria2016} predicted that the eastern limb of HD 209458b should be warmer than the western limb by around 200 K and that NH$_3$ could be enhanced by an order of magnitude on the cooler western limb. The retrievals presented here do not consider such additional effects, with our present ability to disentangle the influence of clouds and chemistry contingent on both a sufficiently long spectral baseline (i.e. optical + near-infrared data) and the planet itself possessing a partially cloudy nature. Ultimately, the constraints derived by a retrieval algorithm are specific to the framework of the assumed models, and it remains to be investigated if differences in limb P-T profiles and chemical abundances can be extracted using current or near-future transmission spectra.

Though future facilities, such as the James Webb Space Telescope (JWST), will undoubtedly revolutionise our understanding of exoplanet atmospheres, so much can still be accomplished with currently available high-precision HST spectra. Our results suggest that the key to obtaining precise chemical abundances from cloudy transmission spectra is rooted in the partially cloudy nature of their terminator; it is the stellar light transiting through the cloud-free region that facilitates breaking many apparent degeneracies between clouds and chemistry. Therefore clouds need not be an insurmountable issue for sufficiently high-precision HST transmission spectra. Now, 8 years after the advent of atmospheric retrieval, the time has come to move beyond one-dimensional models. As we enter the golden age of retrieval, the future is inherently two-dimensional.

\section*{Acknowledgements}

RM would like to acknowledge financial support from the Science and Technology Facilities Council (STFC), UK, towards his doctoral program. RM thanks Siddharth Gandhi for helpful discussions on the forward model. We acknowledge David Sing for making their spectral data publicly available, and the anonymous reviewer for their thoughtful comments on the manuscript.

\bibliographystyle{mnras}
\bibliography{HD209_New_Light} 

%%%%%%%%%%%%%%%%%%%%%%%%%%%%%%%%%%%%%%%%%%%%%%%%%%

%%%%%%%%%%%%%%%%% APPENDICES %%%%%%%%%%%%%%%%%%%%%

\appendix

\section{Derivation of The Two-Dimensional Transit Depth}\label{Appendix:derivation}

For completeness, here we present a concise derivation of the wavelength-dependant transit depth (Equation \ref{eq:transit_depth_two_dim}) for a two-dimensional atmosphere. The transit depth is defined as the fractional stellar flux difference induced as a planet transits its parent star
\begin{equation}
\Delta_{\lambda} = \frac{F_{\lambda,\, \mathrm{out}}-F_{\lambda,\, \mathrm{in}}}{F_{\lambda,\, \mathrm{out}}}
\label{eq:transit_depth}
\end{equation}
where $F_{\lambda, \, \mathrm{out}}$ is the spectral flux observed outside of the transit and $F_{\lambda, \, \mathrm{in}}$ is the spectral flux observed during the transit. The fluxes are in turn defined as integrals of the intensity over solid angle, projected in the observer's line of sight \citep{Seager_book}
\begin{equation}
F_{\lambda} = \int_{\Omega} I_{\lambda} \, \hat{\mathbfit{n}} \cdot \hat{\mathbfit{k}} \,  d\Omega
\label{eq:flux}
\end{equation}
where $\hat{\mathbfit{n}}$ is a unit vector in the direction of propagation of a beam, $\hat{\mathbfit{k}}$ is the unit vector in the direction of the observer and $\Omega$ is the solid angle subtended at the observer.

The flux outside of transit is solely due to the star. From the geometry in Figure \ref{fig:Transit}, we see that $\hat{\mathbfit{n}} \cdot \hat{\mathbfit{k}} = 1$ and $d\Omega= dA/D^2 = b \, d\phi db/D^2$. Thus we have
\begin{equation}
F_{\lambda, \, \mathrm{out}} = \int_{0}^{2\pi} \int_{0}^{R_*} I_{\lambda,\, *} \left(\frac{b}{D^2}\right) db d\phi = \frac{1}{2\pi} \int_{0}^{2\pi} \pi I_{\lambda, \, *} \left(\frac{R_{*}}{D}\right)^2 \, d\phi
\label{eq:F_out}
\end{equation}
where $I_{\lambda,\, *}$ is the intensity at the stellar photosphere. In carrying out the radial integral in the second equality, we have implicitly assumed the stellar intensity is uniform as a function of the impact parameter. This is valid as non-uniform stellar disks, due to limb darkening, are accounted for during data reduction.

The flux during the transit is comprised of three components: flux from the planet itself, stellar flux transmitting through the planetary atmosphere and stellar flux from the disk of the host star surrounding the planet
\begin{equation}
\begin{split}
F_{\lambda, \, \mathrm{in}} = \, & F_{\lambda, \, P} + F_{\lambda, \, \mathrm{A}} + F_{\lambda, \, *} \\
& F_{\lambda, \, P} = \frac{1}{2\pi} \int_{0}^{2\pi} \int_{0}^{R_p} I_{\lambda,P} \left(\frac{2\pi b}{D^2}\right) \, db \, d\phi \\
& F_{\lambda, \, \mathrm{A}} = \frac{1}{2\pi} \int_{0}^{2\pi} \int_{0}^{R_p+ H_A} I_{\lambda,*}(\tau_{\lambda}(b, \phi)) \left(\frac{2\pi b}{D^2}\right) db \, d\phi \\ 
& F_{\lambda, \, *} = \frac{1}{2\pi} \int_{0}^{2\pi} \int_{R_p + H_A}^{R_*} I_{\lambda,*} \left(\frac{2\pi b}{D^2}\right) db \, d\phi
\label{eq:F_in}
\end{split}
\end{equation}
The first integral will evaluate to zero, as the thermal intensity from the planetary nightside, $I_{\lambda,P}$, is negligible compared to the stellar flux. $I_{\lambda,*}(\tau_{\lambda}(b, \phi))$ is the attenuated stellar intensity at impact parameter $b$ and azimuthal angle $\phi$ following passage through a region of the atmosphere with optical depth $\tau_{\lambda}(b, \phi)$. Note that the lower impact parameter limit of the second integral is set to zero to account for the possibility of rays of a given wavelength passing through the planet below $R_p$. Where this may occur is dictated by the local conditions (particularly clouds) at a given azimuthal angle. Regions with low opacity will posses a much deeper opaque radius. Accounting for this requires full evaluation over all impact parameters, without assuming a common lower value for all wavelengths. The third integral is readily evaluated:
\begin{equation}
F_{\lambda, \, *} = \frac{1}{2\pi} \int_{0}^{2\pi} \frac{\pi I_{\lambda,*}}{D^2} \left(R_*^2 - (R_p + H_A)^2 \right) d\phi
\label{eq:3rd_term_F_in}
\end{equation}
To evaluate the second integral, we require an expression for $I_{\lambda,*}(\tau_{\lambda}(b, \phi))$. We make use of the fact that, in the case of transmission of stellar radiation, there will be no emission or scattering into the beam. This renders the solutions of the equation of radiative transfer into simple exponential attenuation, given by Beer's law: $I_{\lambda}(\tau_{\lambda}) = I_{\lambda}(0) \, e^{-\tau_{\lambda}}$. Substituting this into the second integral in \ref{eq:F_in} yields
\begin{equation}
F_{\lambda, \, \mathrm{A}} = \frac{1}{2\pi} \int_{0}^{2\pi} \int_{0}^{R_p + H_A} \frac{2\pi I_{\lambda,*}}{D^2} \left(b \, e^{-\tau_{\lambda}(b, \phi)}\right) db \, d\phi
\label{eq:2nd_term_F_in}
\end{equation}
Substituting this and Equation \ref{eq:3rd_term_F_in} into Equation \ref{eq:F_in} gives
\begin{equation}
\begin{split}
F_{\lambda,\, \mathrm{in}} = \frac{1}{2\pi} \int_{0}^{2\pi} \frac{\pi I_{\lambda,*}}{D^2} \bigg( & R_{*}^2 - R_{p}^2 -2R_p H_A - H_{A}^2 \\
& + 2 \int_{0}^{R_p+ H_A} b e^{-\tau_{\lambda}(b, \phi)} db \bigg) \, d\phi
\label{eq:F_in_2}
\end{split}
\end{equation}
Noting that $2 \int_{R_p}^{R_p + H_A} b \, db = 2R_p H_A + H_{A}^2$ and splitting the integral in Equation \ref{eq:F_in_2} gives
\begin{equation}
\begin{split}
F_{\lambda,\, \mathrm{in}} = \frac{1}{2\pi} \int_{0}^{2\pi} \frac{\pi I_{\lambda,*}}{D^2} \bigg\{ & R_{*}^2 - R_{p}^2 + 2 \int_{0}^{R_p} b e^{-\tau_{\lambda}(b, \phi)} db \\
&  + 2 \int_{R_p}^{R_p+ H_A} b \left(e^{-\tau_{\lambda}(b)} -1\right) db \bigg\} \, d\phi
\label{eq:F_in_3}
\end{split}
\end{equation}

Finally, the flux during and outside transit can be substituted into Equation \ref{eq:transit_depth} to determine the transit depth
\begin{equation}
\Delta_{\lambda} = \frac{1}{2 \pi} \int_{0}^{2 \pi} \delta_{\lambda}\, (\phi) ~ d \phi
\label{eq:Delta}
\end{equation}
where $\delta_{\lambda}\, (\phi)$ is the `one-dimensional' transmission spectrum that would result from assuming an axially symmetric atmosphere with the same properties as the two-dimensional atmosphere possesses at polar angle $\phi$ 
\begin{equation}
\delta_{\lambda} = \frac{R_{p}^2 + 2 \displaystyle\int_{R_p}^{R_p+ H_A} b \left(1 - e^{-\tau_{\lambda}(b, \phi)} \right) db - 2 \displaystyle\int_{0}^{R_p} b e^{-\tau_{\lambda}(b, \phi)} db}{R_{*}^2}
\label{eq:delta}
\end{equation}
As an additional geometric interpretation, we note that this expression for the one-dimensional transit depth is equivalent to taking a perfectly opaque disk of radius $R_p + H_A$ and subtracting the transmitted stellar light. The transmission is simply the integral of the relative area of successive annuli (compared to that of the disk), each weighted by $e^{-\tau_{\lambda}}$.

%%%%%%%%%%%%%%%%%%%%%%%%%%%%%%%%%%%%%%%%%%%%%%%%%%

% Don't change these lines
\bsp	% typesetting comment
\label{lastpage}
\end{document}